\documentclass[epj]{svjour}
\usepackage{psfig}
\begin{document}

\title{Analysis of {\boldmath $\rm \Theta^+$} production in 
{\boldmath ${\rm K^+-Xe}$} collisions}
\author{A. Sibirtsev\inst{^1}, J. Haidenbauer\inst{^1}, S. Krewald\inst{^1}, 
Ulf-G. Mei{\ss}ner\inst{^{1,2}}}

\institute{
Institut f\"ur Kernphysik (Theorie), Forschungszentrum J\"ulich,
D-52425 J\"ulich, Germany
\and
Helmholtz-Institut f\"ur Strahlen- und Kernphysik (Theorie), 
Universit\"at Bonn, Nu\ss allee 14-16, D-53115 Bonn, Germany
}

\date{Received: date / Revised version: date}

\abstract{
The reaction $K^+Xe{\to}K^0pX$ is investigated in a meson-exchange
model including rescattering of the secondary protons with the aim to
analyze the evidence for the $\Theta^+$(1540) resonance reported by the
DIANA collaboration. We confirm that the
kinematical cuts introduced by the DIANA collaboration efficiently suppress
the background to the $K^+n\to K^0p$ reaction which may contribute to the 
$\Theta^+$(1540) production. 
We find that these kinematical cuts do not produce a narrow structure in 
the $K^0p$ effective mass spectra near 1540 MeV. We study the effect of a 
narrow $\Theta^+$ resonance of both positive and negative parity in 
comparison with the DIANA data. 
We show that the $K^+Xe{\to}K^0pX$ calculations without
$\Theta^+$ contribution as well as the results obtained with a
$\Theta^+$ width of 1~MeV are in comparably good agreement with 
the DIANA results. More dedicated experiments are called for to
establish this exotic baryon resonance.
}
\maketitle

\section{Introduction}

The evidence for a narrow $S{=}+1$ baryon resonance in photoproduction
from the neutron reported in Ref. \cite{Nakano} and the subsequent
observation of such a structure in the invariant
mass spectrum of the kaon-nucleon ($KN$) system in other reactions 
led to a wealth of theoretical investigations dealing with this 
topic. By now the production of this exotic baryon state,
called $\Theta^+$(1540), in basically every elementary
photo-, kaon- or nucleon induced process has been worked out in
detail assuming positive as well as negative parity for the
$\Theta^+$ state. It is interesting, however, that so far -- to
the best of our knowledge -- there is not a single calculation
that tries to provide a theoretical description of any of those
experiments where the $\Theta^+$ baryon resonance was actually detected. 
Certainly, one has to concede that in several of the works
where evidence for the $\Theta^+$ was claimed, cuts are
applied which are not described in very detail and therefore 
a direct and quantitative comparison of any model calculation
with those data would be difficult if not impossible. Luckily, there
are also exceptions and one of them is the result of the
DIANA collaboration \cite{Dolgolenko}, where the 
$\Theta^+$ resonance was observed in $K^+$ collisions with
$Xe$ nuclei. 

In this paper we present a detailed and microscopic
calculation of the $K^0p$ effective mass spectrum for the
charge-exchange reaction $K^+Xe{\to}K^0pX$. We aim at a
quantitative analysis of the mass spectrum that was published
by the DIANA collaboration \cite{Dolgolenko}. The main ingredient of 
our calculation is an elementary $KN$ scattering amplitude
that is taken from the meson-exchange model developed by
the J\"ulich group \cite{Juel1,Juel2}. The role and
significance of the $\Theta^+$(1540) for a quantitative
description of the DIANA data is investigated by 
employing variants of the J\"ulich $KN$ model where a
$\Theta^+$ resonance with different widths was included
consistently \cite{Haidenbauer1,Sibirtsev0}. Since the
parity of the $\Theta^+$ resonance is not yet known we 
consider here the cases of positive as well as negative
parity. We assume this exotic resonance to have spin 1/2
throughout. 

The paper is structured in the following way. 
In the subsequent section we provide a short description
of the $KN$ interaction model of the J\"ulich group. 
We also explain how the $\Theta^+$ resonance is 
implemented into the model. Furthermore we point to
the importance of unitarity and emphasize the 
interplay between the background phase, provided by 
the non-pole part of the interaction, and the resonance
structure. This feature plays a decisive role
in our analysis of the DIANA data. 
In section 3 first we describe how the calculations of the
$K^0p$ effective mass spectrum for the
charge-exchange reaction $K^+Xe{\to}K^0pX$ 
are performed. Then we present a detailed quantitative
comparison of our model calculation with the data 
of the DIANA collaboration, considering their results
before cuts were applied as well as those with cuts. 
The paper ends with a short summary.

\section{The kaon-nucleon amplitude}

We use the J\"ulich meson-exchange model for the $KN$ 
interaction.  A detailed description of this model is given in 
Refs.~\cite{Juel1,Juel2}. It was constructed along
the lines of the full Bonn $NN$ model~\cite{MHE} and its extension 
to the hyperon-nucleon $YN$ system~\cite{Holz}. 
Specifically, this means that one
has used the same scheme as a time-ordered perturbation theory
and the same type of processes. Moreover, most of the vertex parameters 
(coupling constants and cut-off masses of the vertex form-factors)
appearing in the diagrams that contribute to the interaction
potential $V$ have been fixed already by the study of other reactions. 

Once the $KN$ interaction potential $V$ is derived, the corresponding 
reaction amplitude $T$ is then obtained by solving a Lippmann-Schwinger 
equation defined by time-ordered perturbation theory,
\begin{equation}
T = V + V G_0 T \ ,
\label{LSE}
\end{equation}
where $G_0$ is the free propagator.
Taking the solution of such a scattering equation implies that 
the resulting reaction amplitude $T$ automatically fulfils the
requirements of (two-body) unitarity. 
In the present investigation we use the $KN$ model (I) described 
in Ref.~\cite{Juel2}. Results for phase shifts and also for 
differential cross sections and polarizations can be 
found in Ref.~\cite{Juel2}.  
Evidently this model yields a good overall reproduction of all
presently available empirical information on $KN$ scattering. 
Specifically, it describes the data up to kaon laboratory momenta of 
$p_K{\approx}$1 GeV/c, i.e. well beyond the region of the observed 
$\Theta^+$ resonance structure which corresponds to the momentum 
$p_K{\simeq}$440 MeV/c.
 
The $\Theta^+$ resonance is included in the model by 
adding a pole diagram to the other diagrams that contribute to $V$
with a bare mass ${\hat M}_{\Theta^+}$ and a bare coupling constant
${\hat g}_{KN\Theta^+}$ 
When this interaction is then iterated in the Lippmann-Schwinger 
equation the ${KN\Theta^+}$ vertex gets dressed by the non-pole part 
of the interaction and the $\Theta^+$ acquires a width and also 
its physical mass via self-energy loops. Note that the bare mass and 
bare coupling constant are free parameters that are adjusted to 
obtain the desired physical mass and width of the $\Theta^+$ resonance
which are ${M}_{\Theta^+}$=1540 MeV and 
${\Gamma}_{\Theta^+}$=1, 5, and 10 MeV. 

\begin{figure}[t]
\vspace*{3mm}
\hspace*{2.mm}\psfig{file=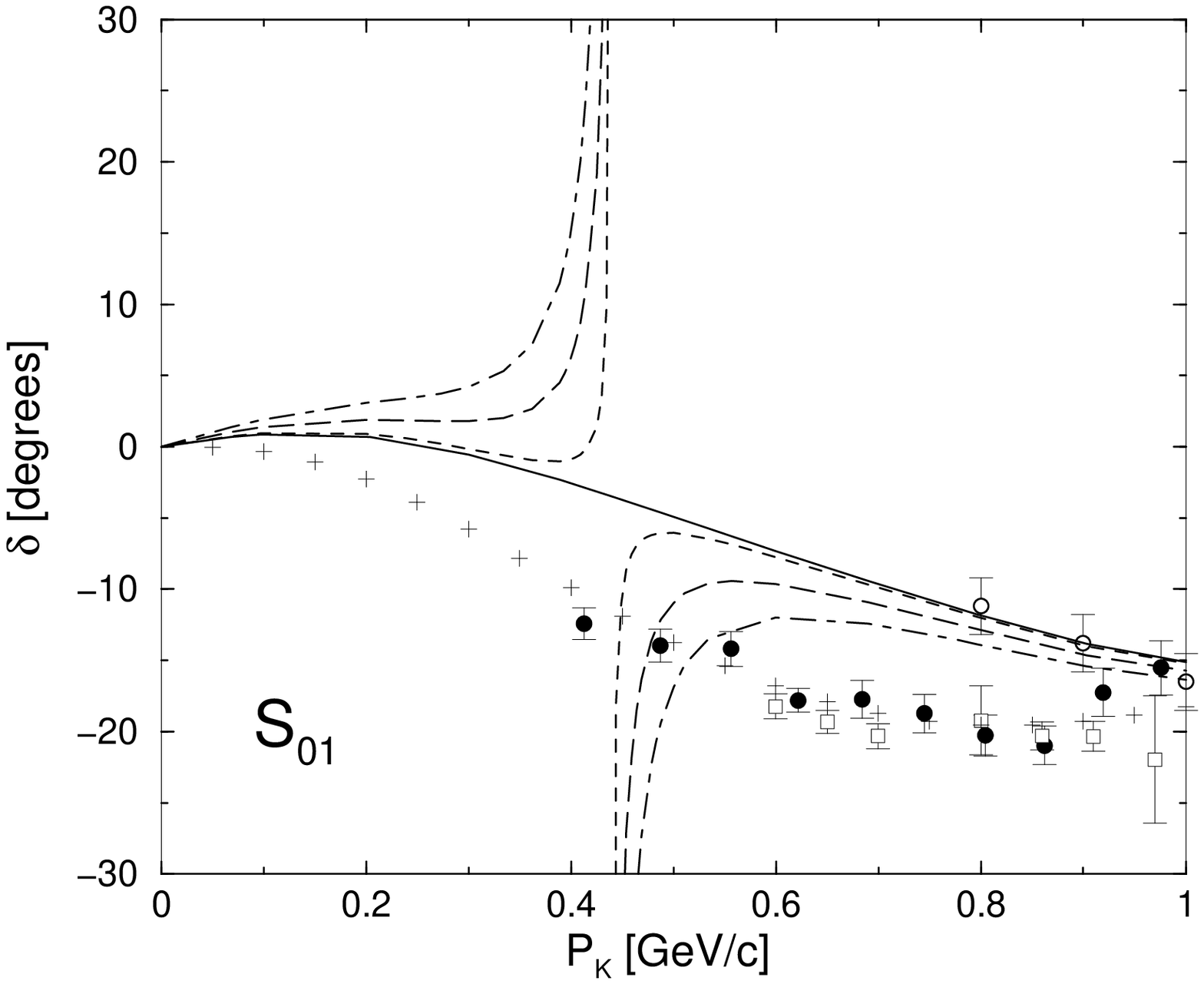,width=8.0cm,height=7.5cm}
\hspace*{2.mm}\psfig{file=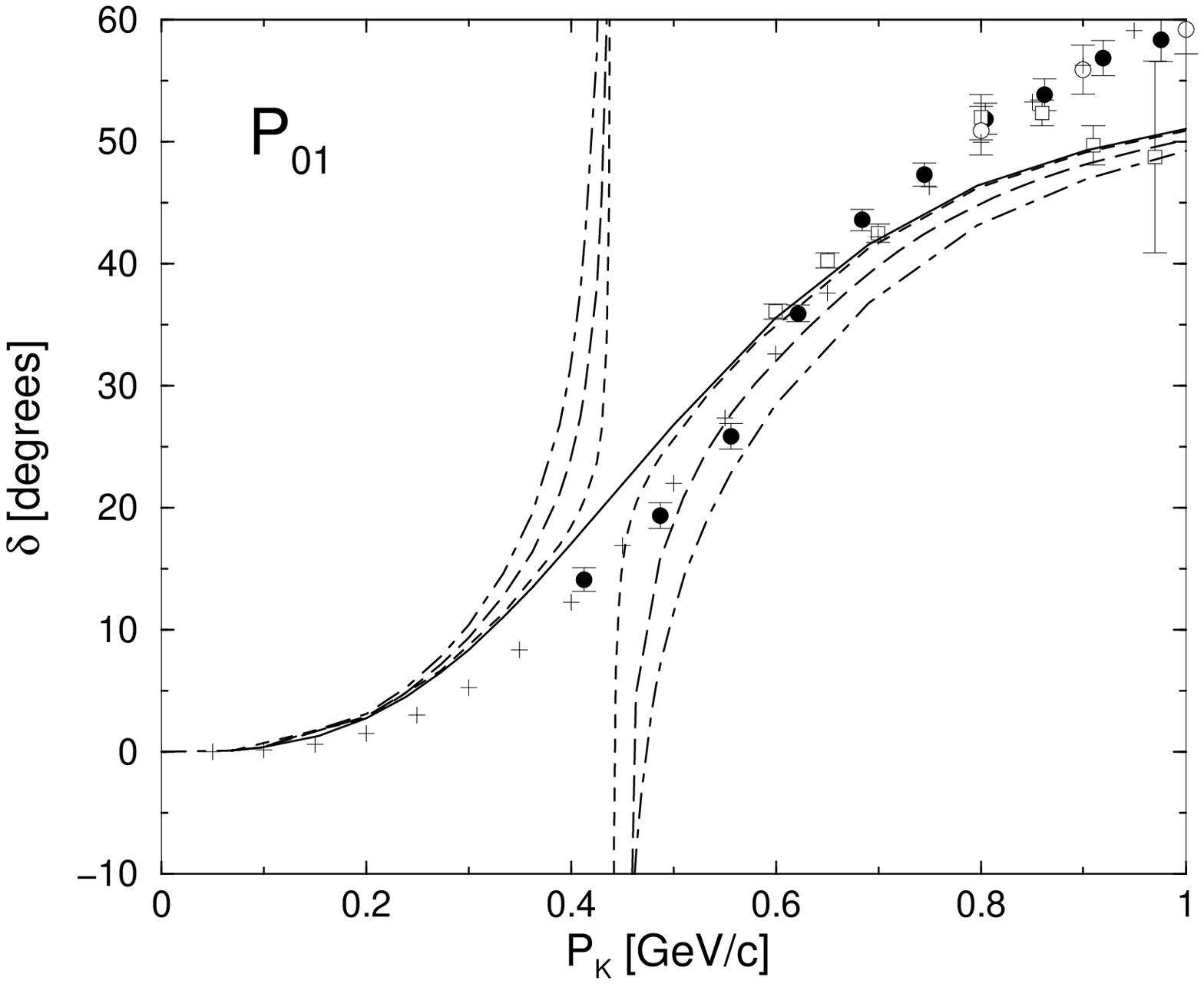,width=8.0cm,height=7.5cm}
\vspace*{-1mm}
\caption{The $S_{01}$ and $P_{01}$ phase shifts for $KN$ scattering as a
function of the kaon momentum. The solid lines show the calculations 
without $\Theta^+$ contribution. The short-dashed line is our
calculation with $\Gamma_{\Theta^+}$=1 MeV, the
long-dashed line corresponds to $\Gamma_{\Theta^+}$=5 MeV
and the dash-dotted line to $\Gamma_{\Theta^+}$=10 MeV. Note that
the phases for the resonance cases are shown modulo $\pi$.}
\label{diana8}
\end{figure}

\begin{figure}[t]
\vspace*{3mm}
\hspace*{2.mm}\psfig{file=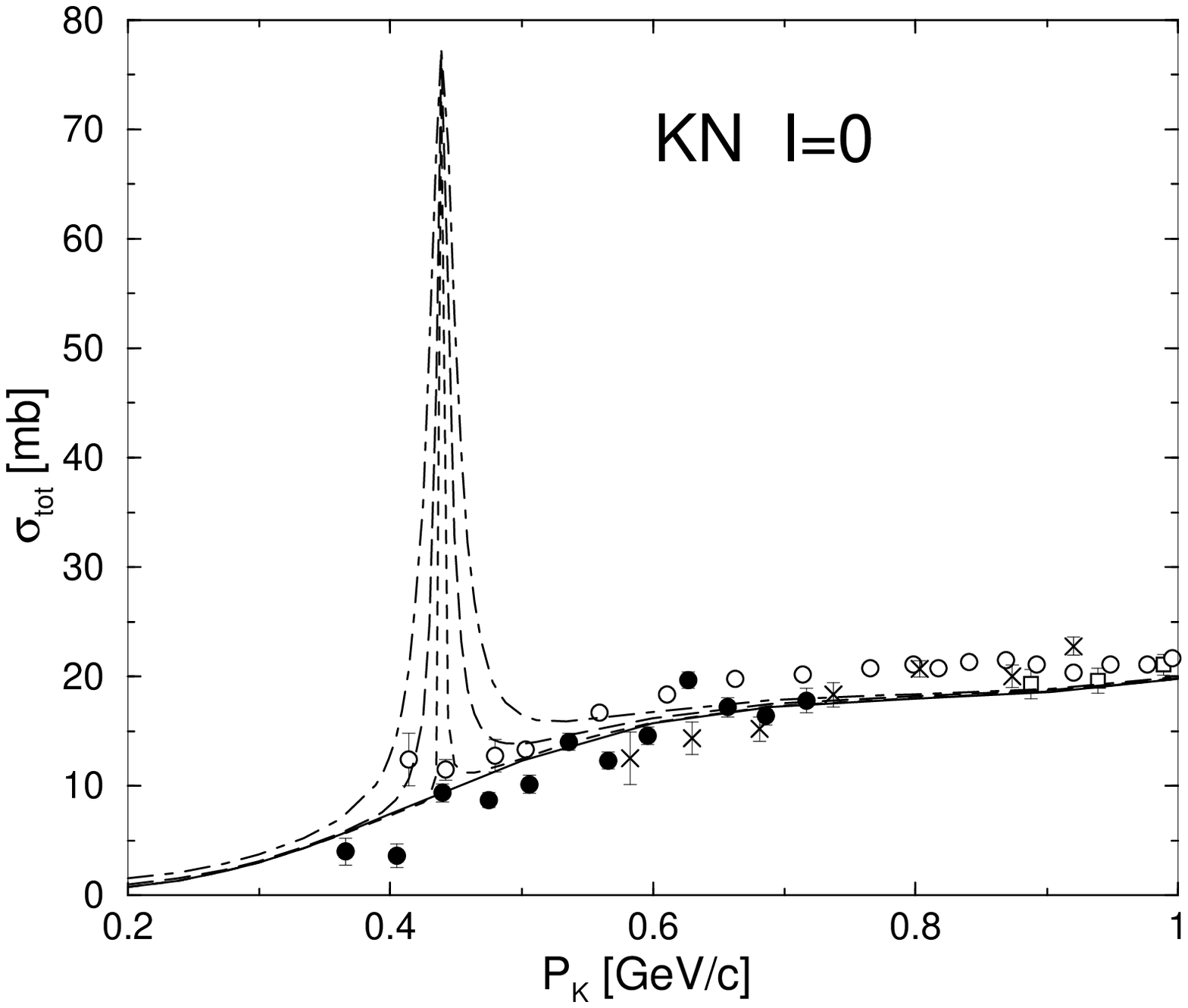,width=8.0cm,height=8.1cm}
\hspace*{2.mm}\psfig{file=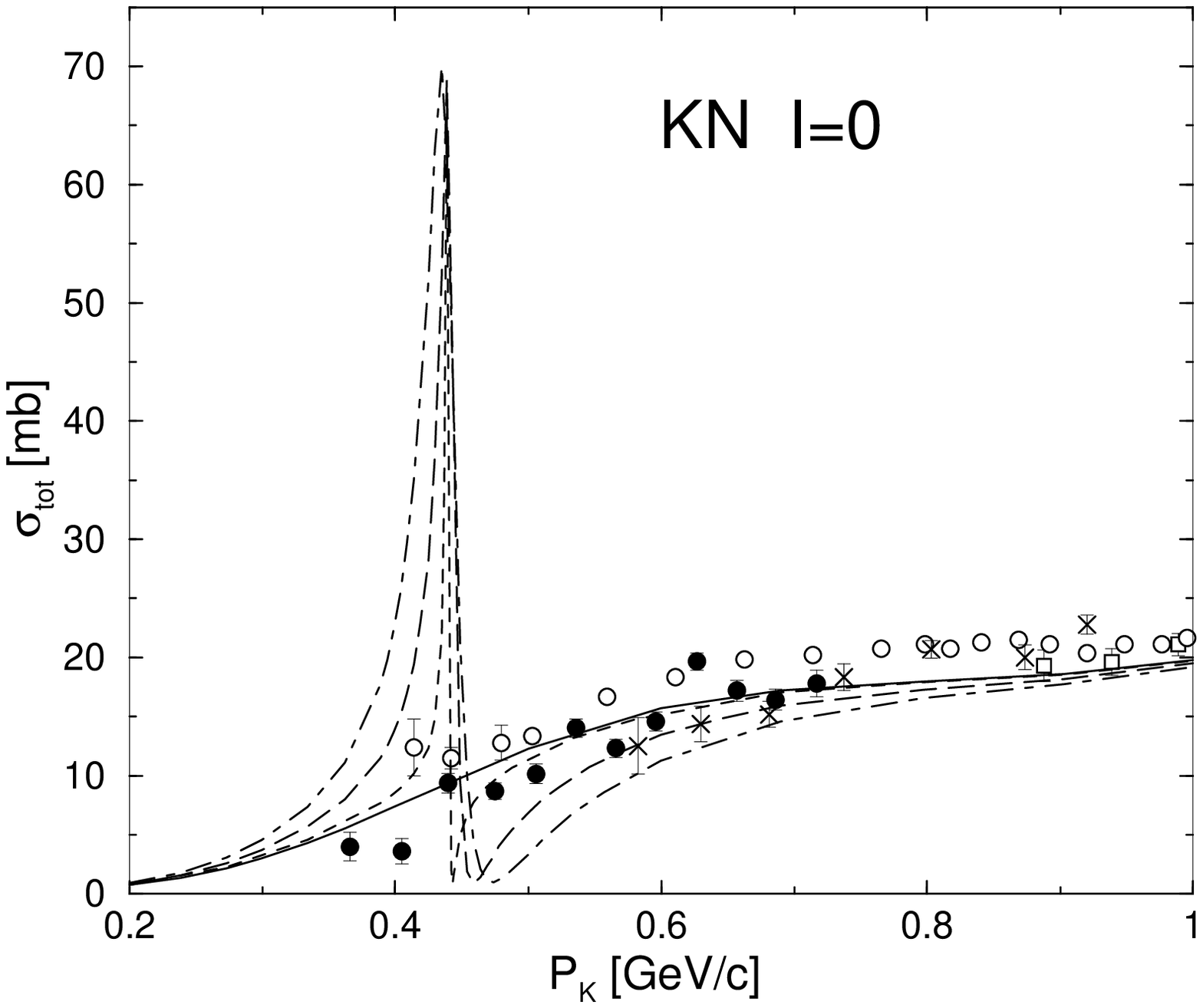,width=8.0cm,height=8.1cm}
\vspace*{-1mm}
\caption{The $KN$ reaction cross section for the $I{=}0$ channel
as a function of the kaon momentum. Results are shown for
negative (top) and positive (bottom) parity of the $\Theta^+$
resonance. The various curves correspond to different $\Theta^+$ 
widths,  
$\Gamma_{\Theta^+}$=1~MeV (short dashed), 
5~MeV (long dashed), and 10~MeV (dash-dotted),
while the solid line is the prediction without a $\Theta^+$ 
contribution.
Data are taken from Refs.~\cite{Bow70} (filled circles), 
\cite{Coo70} (squares), \cite{Car73} (open circles), and
\cite{Bow73} (crosses). 
}
\label{knsig}
\end{figure}

Depending on the parity of the ${\Theta^+}$ the resonance 
will contribute to different partial waves of the $KN$ interaction. 
If the ${\Theta^+}$ is a $J^P$=${1\over 2}^+$ state,
as predicted by e.g. the chiral quark-soliton model,
then it couples to the $P_{01}$ channel (we use the standard
spectroscopic notation $L_{I \,2J}$). If $\Theta$ has the opposite parity
it will occur in the $S_{01}$ partial wave. These two phases 
exhibit quite different characteristics as can be seen in Fig.
\ref{diana8}. The $S_{01}$ phase shift is small and basically repulsive. In
fact $\delta_{S_{01}}$ is practically zero up to the momentum of 
$p_K$=440 MeV/c corresponding to the ${\Theta^+}$ 
resonance mass, cf.  Fig.~\ref{knsig}. 
In contrast the $P_{01}$ phase shift is large and 
attractive. As a consequence, the total $KN$ cross section for 
isospin $I$=0
is dominated by the $P_{01}$ partial wave while the contribution
of the $S_{01}$ partial wave is roughly given by the result on the
very left side of Fig.~\ref{knsig}. 

Adding a resonance in these two partial waves generates again
quite different characteristics. In the $S_{01}$ partial wave
we obtain basically a Breit-Wigner type structure because the
non-resonant background is rather small. In the $P_{01}$ partial 
wave, on the other hand, there is an interference of the
background phase with the added resonance structure. In particular, 
the phase will not only rise beyond $90^0$ (because of the resonance)
but even beyond $180^0$ (because of the large background). For
momenta around the region where the phase shift passes through $180^0$ 
there will be practically no contribution of this partial wave to the 
total cross section because 
$\sigma_{P_{01}}{\propto}\sin^2(\delta_{P_{01}})$. 
As a consequence, a resonance in the
$P_{01}$ partial wave causes a rather striking bump-dip structure in
the $KN$ $I{=}0$ cross section, cf. Fig. \ref{knsig}.

\begin{figure}[t]
\vspace*{3mm}
\hspace*{2.mm}\psfig{file=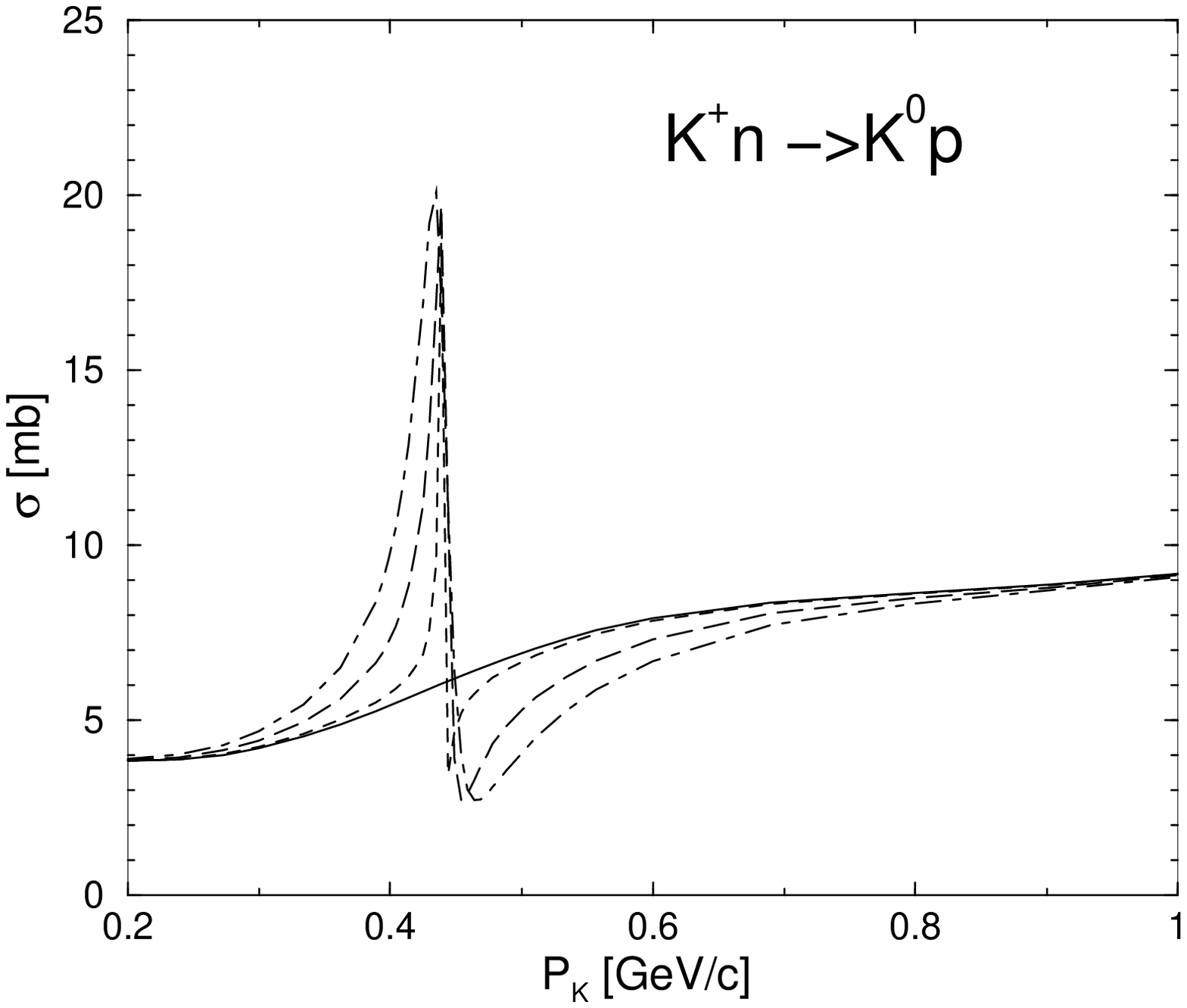,width=8.0cm,height=8.1cm}
\hspace*{2.mm}\psfig{file=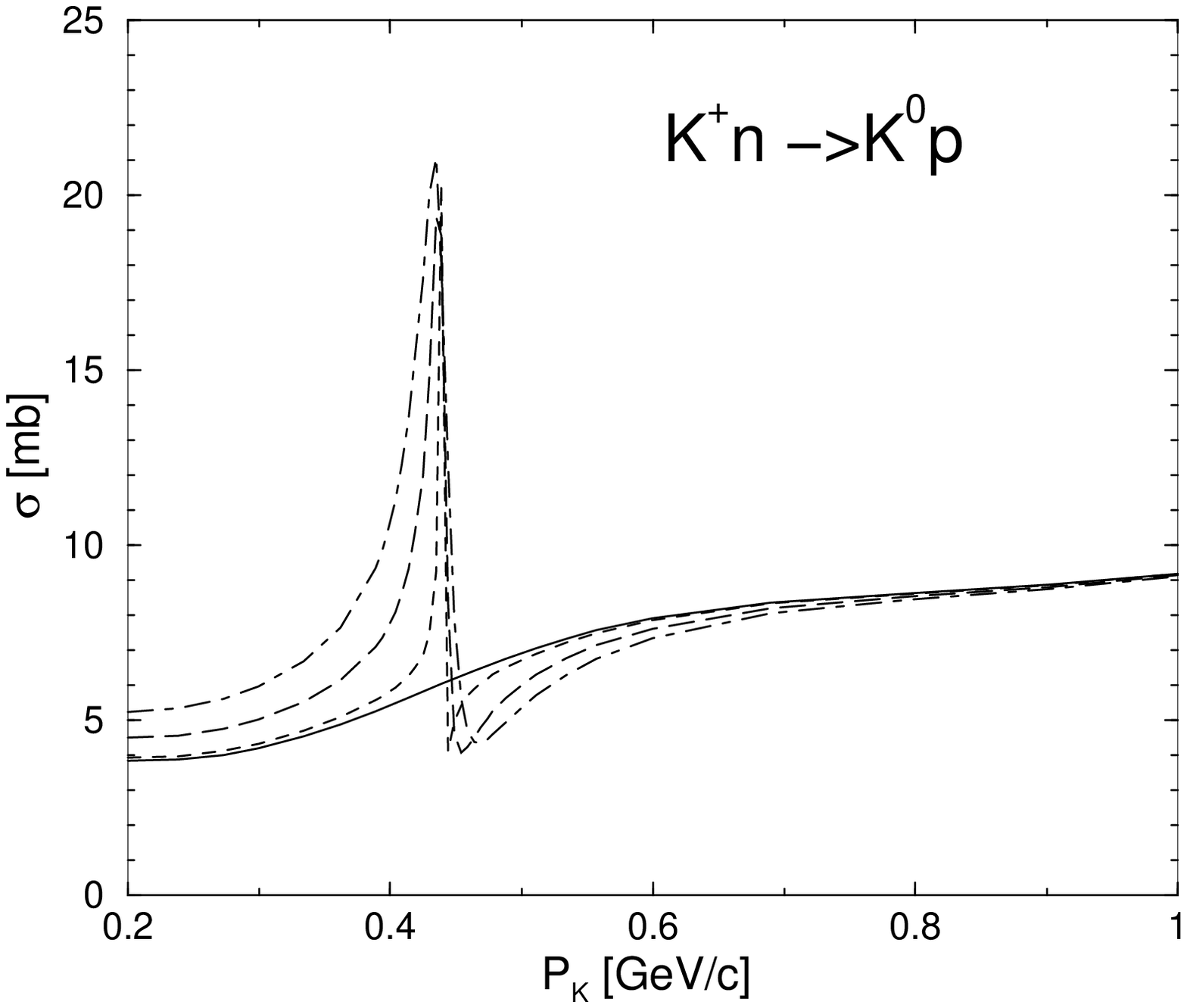,width=8.0cm,height=8.1cm}
\vspace*{-1mm}
\caption{The $K^+n{\to}K^0p$ reaction cross section as a function
of the kaon momentum. For notation, see Fig.~\ref{knsig}.}
\label{diana7}
\end{figure}

Fig.~\ref{diana7} shows the cross section for the charge-exchange
reaction $K^+p{\to}K^0p$ as a function of the kaon momentum. This 
is the elementary process that is needed for analyzing the 
reaction $K^+Xe{\to}K^0pX$ measured by the DIANA collaboration.
Obviously, here the characteristic differences between the $S_{01}$ 
and $P_{01}$ resonances have basically disappeared. Since the 
charge-exchange amplitude is given by the difference of the isospin $I{=}1$ 
and $I{=}0$ amplitudes we now get an interference of the (larger) $S_{11}$
partial wave with the resonant $S_{01}$ partial wave and that leads
to a similar bump-dip structure in the cross section as produced
by the interference of the resonance with the background in the 
$P_{01}$ case. Still one can see that for the negative parity case
the dip after the resonance energy is somewhat less pronounced,
cf. the two cases in Fig.~\ref{diana7}.

This behaviour of the elementary $KN$ and $K^+p{\to}K^0p$ cross
sections should be also reflected in
all of those reactions where the $\Theta^+$ was observed because
there again  the complete $KN$ amplitude enters.
In the next section we examine in detail the influence of these
characteristic features induced by the presence of a $\Theta^+$ 
resonance on the $K^0p$ mass spectrum for the reaction 
$K^+Xe \to K^0pX$ measured by the DIANA collaboration.

We would like to mention that there are no data on the
elementary $K^+ n{\to} K^0 p$ process. However, the charge-exchange
channel has been measured on a deuteron target, i.e. in the
reaction $K^+d{\to}K^0pp$. In Ref. \cite{Sibirtsev0} we have 
investigated this reaction employing the J\"ulich $KN$ model (I),
that is used also in the present work, and we have found that the model 
predictions are in very good agreement with corresponding data. 

\section{The $K^+\rm{Xe}{\to}K^0p\rm{X}$ model.}
\begin{figure}[b]
\vspace*{-8mm}
\hspace*{-0.5mm}\psfig{file=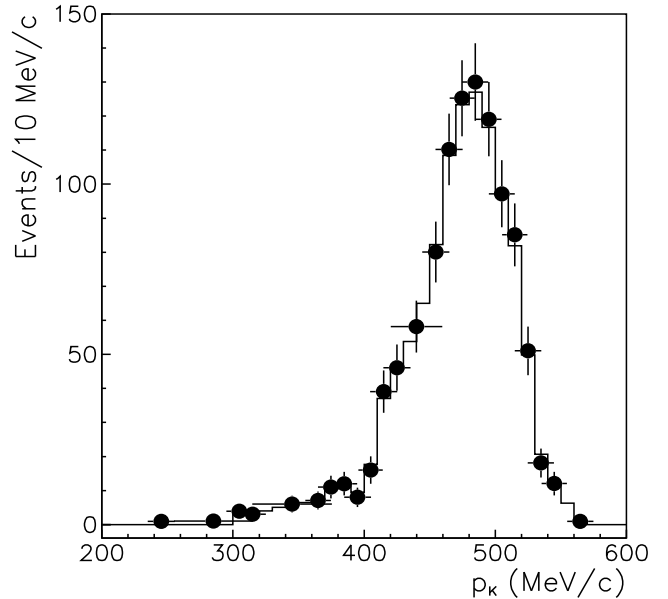,width=9.3cm,height=9.cm}
\vspace*{-8mm}
\caption{The momentum spectrum of the incident $K^+$ mesons. The circles 
show experimental results, while the histogramm is our simulation.}
\label{diana1}
\end{figure}

There are some special details of the DIANA experiment that 
substantially constrain our model calculations. The incident 
$K^+$ mesons have no fixed momentum but are distributed between 
(roughly) 300 and 550~MeV/c as is shown by Fig.~\ref{diana1}. Because
of the reaction kinematics kaons with that spectrum would produce 
a $K^0p$ mass distribution, which is obviously enhanced at large 
masses. In addition the free $K^+n{\to}K^0p$ reaction cross section  
increases by a factor 2.5 within the range 300${\le}p_K{\le}$550~MeV/c, 
as is illustrated in Fig. \ref{diana7}. Therefore, it is natural to expect 
that the observed $K^0p$ mass spectrum should be shifted towards the 
maximal available $K^0p$ mass. 

\begin{figure}[t]
\vspace*{-6mm}
\hspace*{-0.5mm}\psfig{file=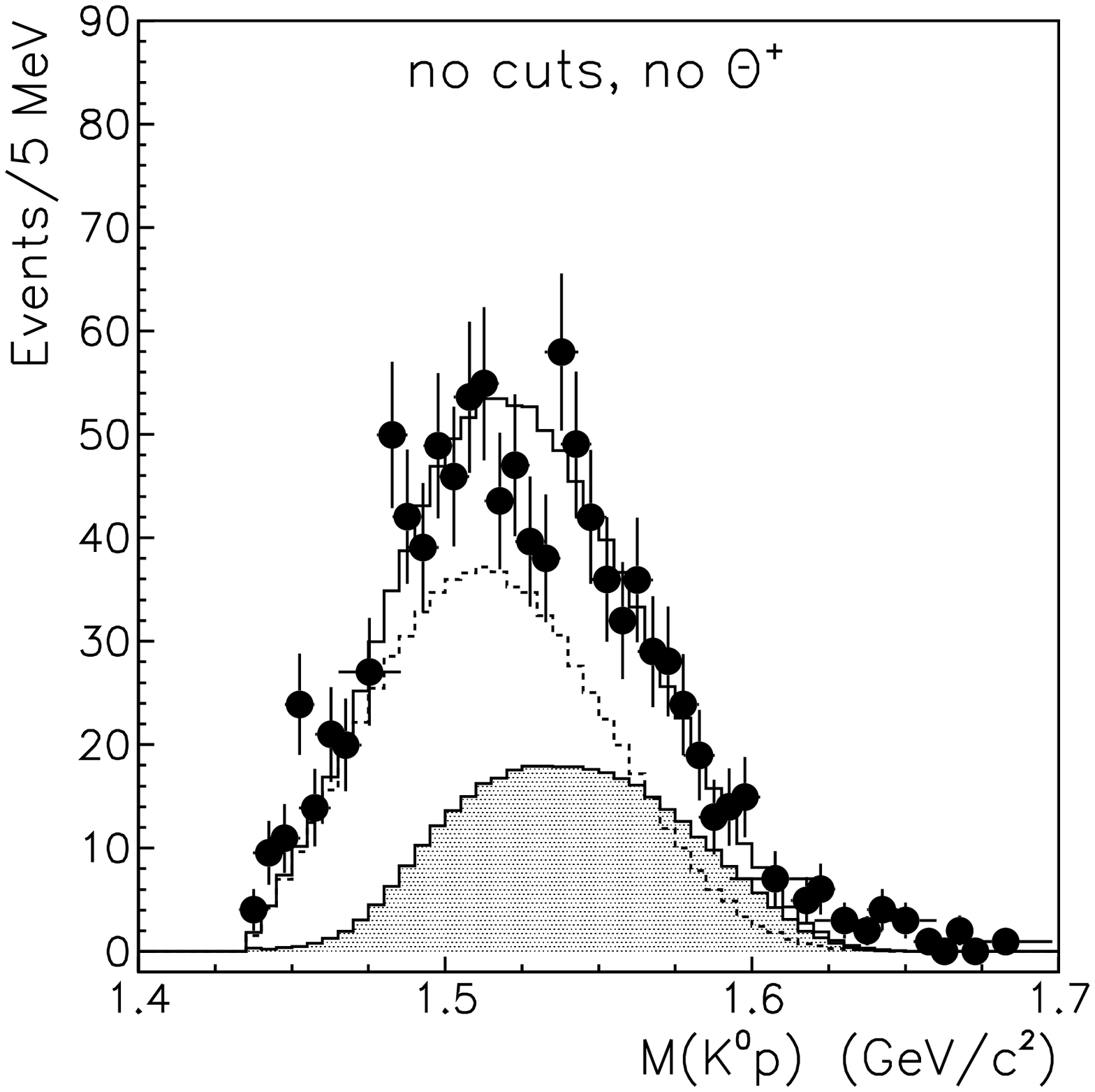,width=9.3cm,height=9.cm}
\vspace*{-12mm}

\hspace*{-0.5mm}\psfig{file=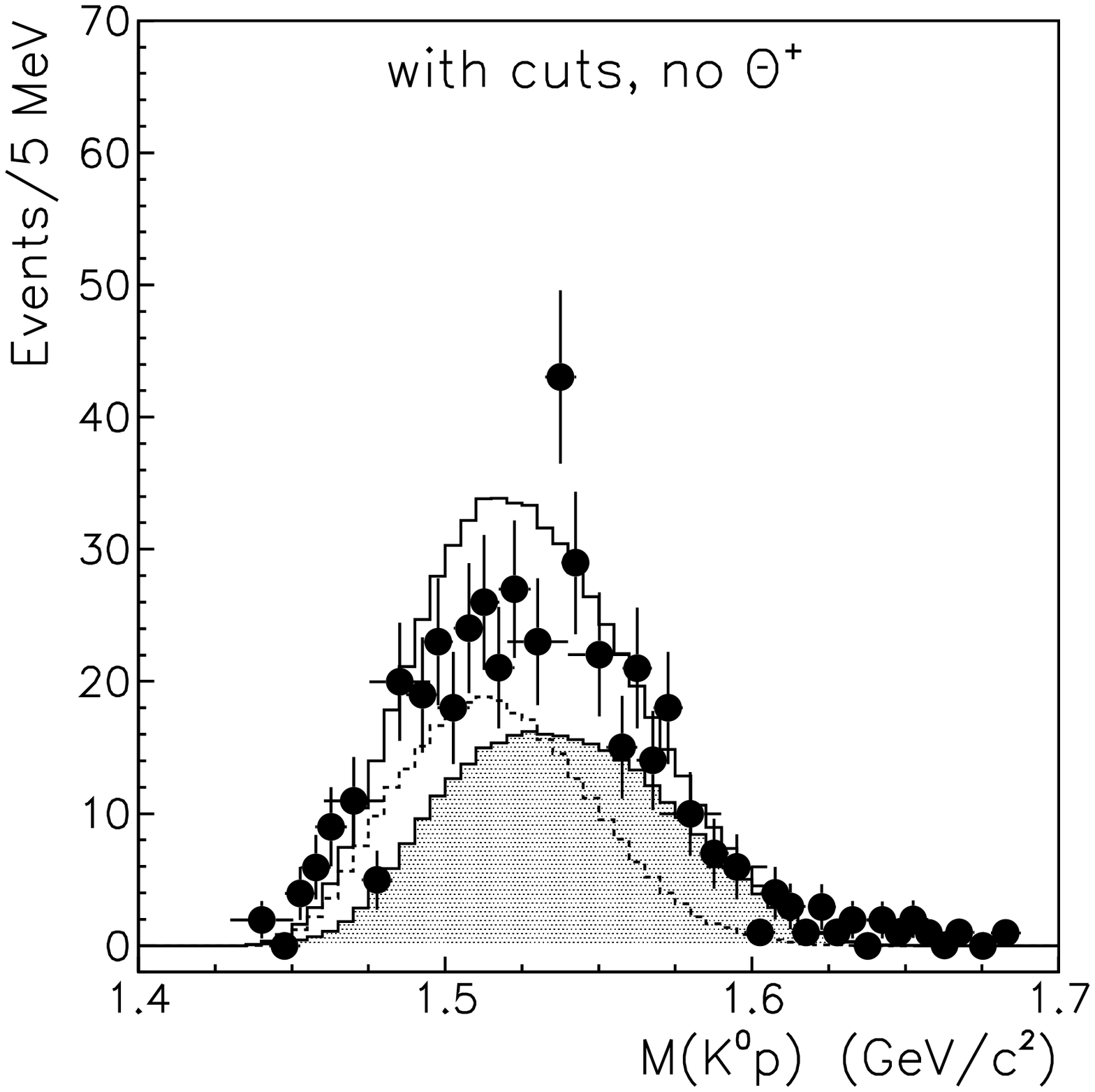,width=9.3cm,height=9.cm}
\vspace*{-8mm}
\caption{The $K^0p$ invariant mass spectrum from the $K^+Xe$ reaction. The 
circles show experimental results from the DIANA 
collaboration~\cite{Dolgolenko}
without (top) and with kinematical cuts (bottom). The histograms are 
our results without the inclusion of a $\Theta^+$ resonance, 
with and without experimental cuts. Hatched histograms 
show the spectrum from direct $K^+n{\to}K^0p$ production on a bound target
neutron, for the dashed curve an additional single rescattering of 
the proton was taken into account, while the solid curve is their sum.}
\label{diana2}
\end{figure}

The measured $K^0p$ mass spectrum
shows an absolutely different shape as is indicated by 
Fig.~\ref{diana2}. Prior to kinematical cuts the 
DIANA data show an enhancement at low $K^0p$ masses. This can
be understood in terms of nuclear effects. As will be shown 
later the enhancement for low masses comes from the nuclear rescattering 
of the final protons or $K^0$ mesons. Through the rescattering in the
nucleus the particle transfers part of its initial momentum to the 
target nucleon by an elastic or inelastic scattering.
Therefore one should model not only the $K^+n{\to}K^0p$  reaction in 
$Xe$, but also the rescattering of the proton and the kaon.

It is clear that the  reaction $K^+N{\to}K^+N$
with $N$ being either a bound proton or neutron can also contribute.
Then the final $K^+$ might interact with a neutron to be converted to a
$K^0$. However, the kaon mean free path of $\lambda_K{\simeq}$5~fm is quite 
large and thus the probability of kaon rescattering in the final
nucleus is smaller as compared to the rescattering of the final 
proton which has $\lambda_p{\simeq}$1.5~fm. Here we do not consider 
the final state interaction of the kaon.

Another feature of the DIANA experiment is the 
availability of the measured $K^0p$ mass spectrum prior and after 
the kinematical cuts. These cuts were used for attenuation of the 
background from nuclear rescattering. In our calculations we are 
comparing to the data with and without kinematical cuts in order to
control the contribution from the direct $K^+n{\to}K^0p$ reaction
in the $Xe$ nucleus and the rescattering processes. 

We evaluate the differential $K^+n{\to}K^0p$ cross section on
a bound neutron from 
\begin{eqnarray}
\frac{d^3\sigma}{dp d\Omega}{=}\!\!\!
\int \!\!\Phi (p_K) dp_K \!\!\!
\int \!\!P(E_q,q) |T_{KN}(s_{KN},\Omega)|^2 dE_qd^3q,\,\,\,\,\,\,\,\,
\label{react1}
\end{eqnarray}
where $\Phi (p_K)$ is the $K^+$-meson spectrum shown in 
Fig.~\ref{diana1} and $T_{KN}$ is the amplitude
for the reaction $K^+n{\to}K^0p$, which depends on the $K^0$-meson 
production angle $\Omega$ and the squared invariant energy of 
the incident kaon and target neutron given by
\begin{eqnarray}
s_{KN}=(E_K+E_q)^2-(\vec{p}_K+\vec{q})^2.
\label{skn}
\end{eqnarray} 
Note that to evaluate Eq.~(\ref{react1}) we utilize the on-shell
amplitude, which should be a reasonable approximation
at low kaon momenta.

The nuclear spectral function  $P(E_q,q)$ defines the joint probability
to find in a nucleus a nucleon with momentum $q$ and total energy $E_q$.
It is clear that the relation between $q$ and $E_q$ in nuclei is not 
given by the free dispersion relation. Within the Thomas-Fermi approximation 
the spectral function is given as
\begin{eqnarray}
P(E_q,q)= \frac{3} {4 \pi q_F^{3/2}} 
\Theta (q_F- |{\bf q}|) \nonumber \\
\times \delta\biggl(E_q-\sqrt{q^2+m_N^2}- U(q,\rho)\biggr),
\end{eqnarray}
where $q_F$ is the Fermi momentum, $m_N$ is the free nucleon mass and the
nuclear potential $U$ depends on the nucleon momentum $q$ and on the density 
$\rho$ as~\cite{Gale,Welke}
\begin{eqnarray}
U(q,\rho) = A_1\frac{\rho}{\rho_0} + 
B_1\left(\frac{\rho}{\rho_0}\right)^{1.24} +
\frac{2 \pi C_1 \Lambda_1}{\rho_0} \nonumber \\
\times \left[
\frac {q_F^2 + \Lambda_1^2 -q^2} {2 q \Lambda_1} \times
\ln \frac {(q+q_F)^2 +\Lambda_1^2} {(q-q_F)^2 +\Lambda_1^2}
\right. \nonumber \\ \left. + \frac {2q_F } {\Lambda_1 }-
2 \left( \arctan \frac {q+q_F} {\Lambda_1} 
- \arctan \frac {q-q_F} {\Lambda_1} \right) \right],
\label{pwelke}
\end{eqnarray}
where $A_1$=--110.44~MeV, $B_1$=140.9~MeV, $C_1$=-64.95~MeV 
and $\Lambda_1$=1.58$q_F$. Furthermore, for the calculations with the
uncorrelated part of the momentum distribution for the finite
nuclei we use $q_F$=220~MeV/c~\cite{Frullani,Atti}.

We neglect the high momentum or correlated part of the spectral 
function, which contributes to subthreshold particle 
production~\cite{Sibirtsev1,Sibirtsev2}, 
but does not affect the results of the present study. 
Indeed the maximum of the spectral function 
for high momenta, which is attributed to 
short-range and tensor correlations, corresponds to a total
energy of the bound nucleon given as~\cite{Sibirtsev1,Atti},
\begin{eqnarray}
E_q\simeq m_N-2\epsilon-\frac{q^2}{2m_N},
\label{eq}
\end{eqnarray}
with $\epsilon{\simeq}7$~MeV. By inserting this energy $E_q$ into 
Eq.~(\ref{skn}) one may show that the high momentum component of
the spectral function would not distort noticeably the
$K^0p$ invariant mass spectrum. 

Note that for the total nuclear momentum distribution the Fermi 
momentum for $^{131}Xe$ nucleus is  
$q_F{\simeq}$265~MeV/c and the corresponding average nucleon 
separation energy is roughly --42~MeV~\cite{Frullani}, which is 
substantially beyond the applicability of the Thomas-Fermi 
approximation. In order to include into the calculations the 
high-momentum tail of the nuclear momentum distribution one needs to
use a realistic spectral function or to explore the dispersion
relation given by Eq.~(\ref{eq}).

However, as was shown in Ref.~\cite{Sibirtsev2} calculations 
within the local Thomas-Fermi approximation are in reasonable agreement 
with the results obtained with realistic spectral functions
evaluated within the orthogonal correlated basis 
approach~\cite{Benhar,Sick} unless the calculations are performed for 
the production of particles at energies substantially below the threshold
in free space. 

\begin{figure*}
\begin{tabular}{cc}
\vspace*{-8mm}\psfig{file=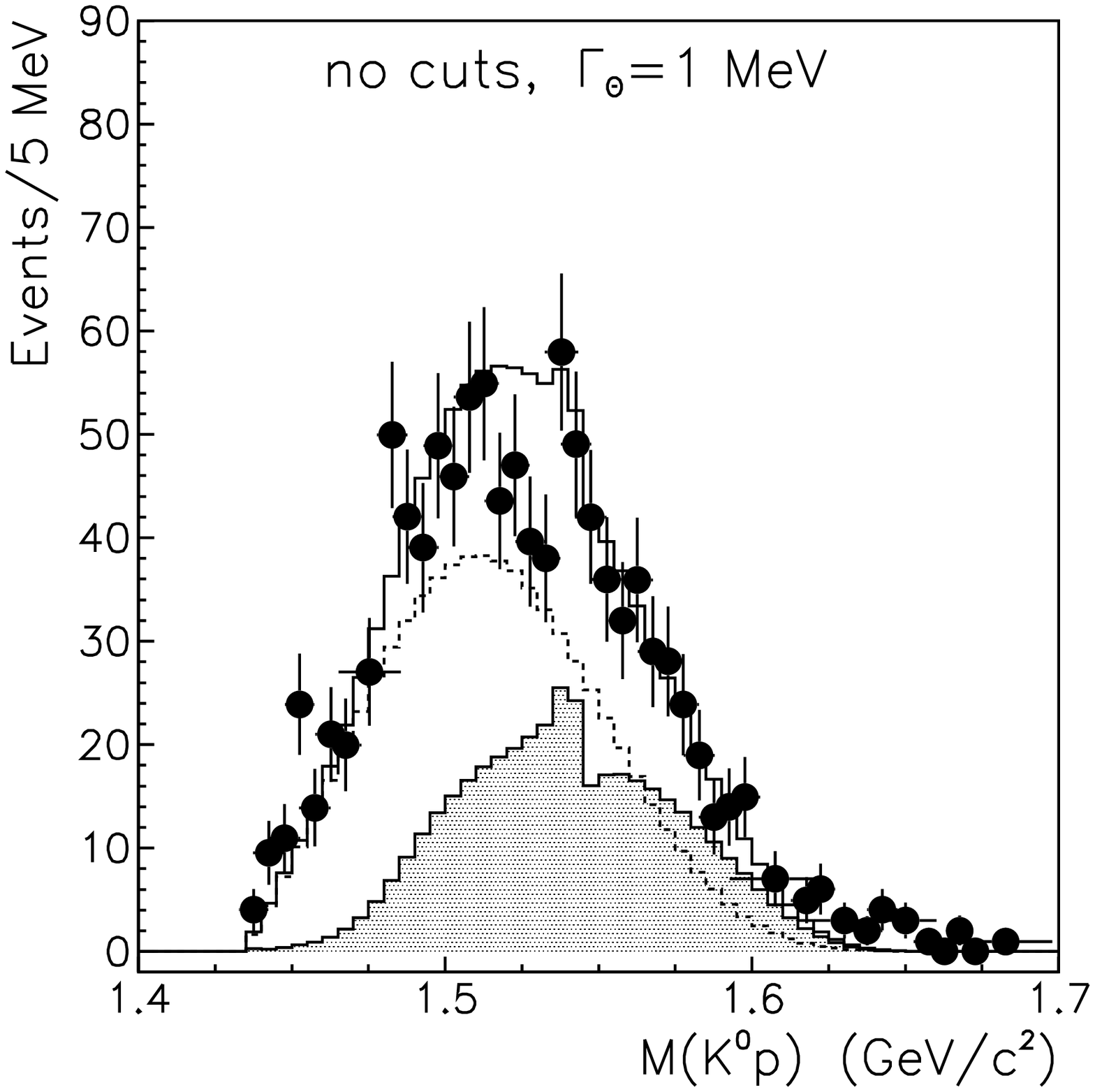,width=9.3cm,height=8.3cm} &
\hspace*{-10mm}\psfig{file=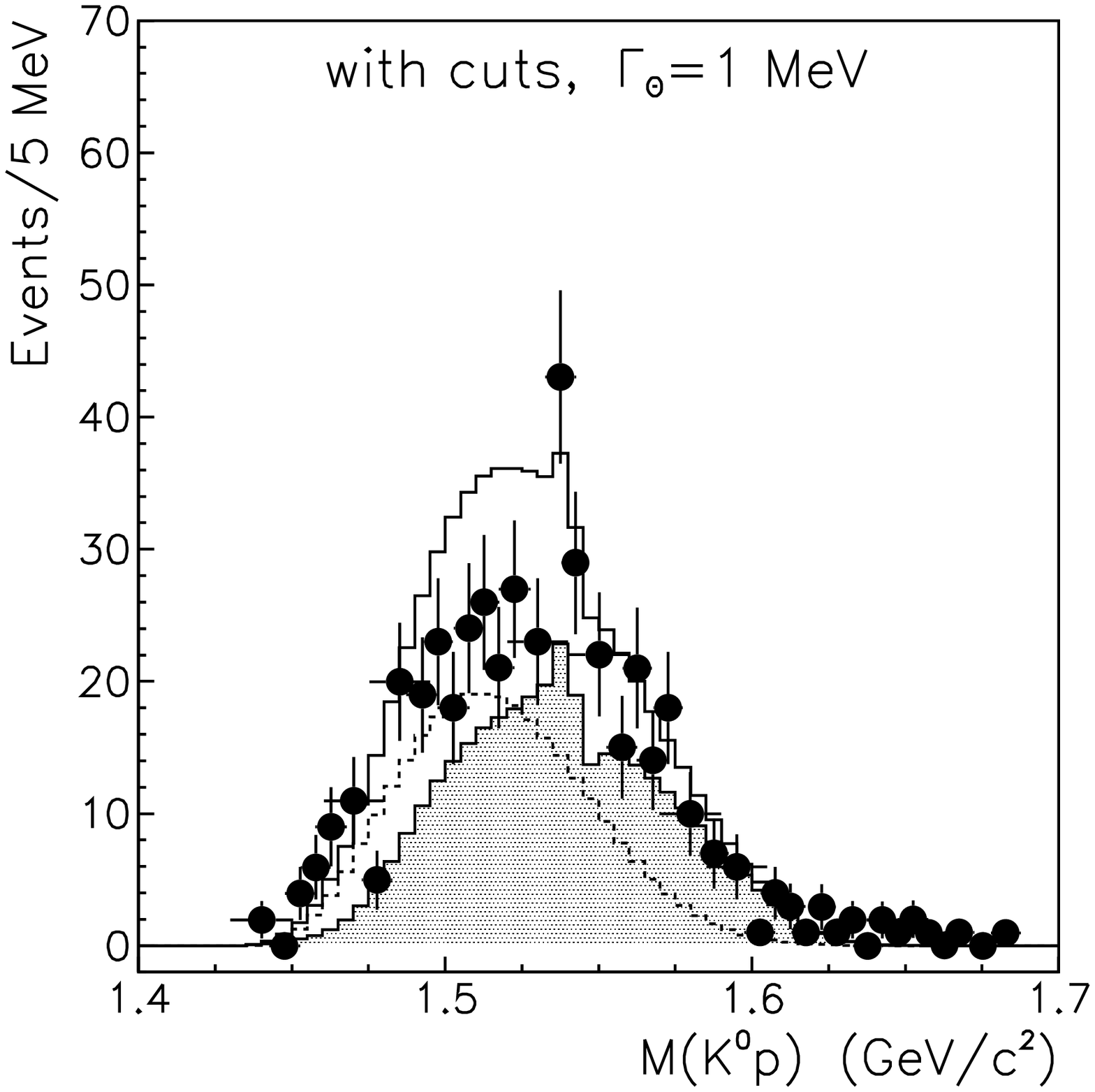,width=9.3cm,height=8.3cm} \\
\vspace*{-8mm}\psfig{file=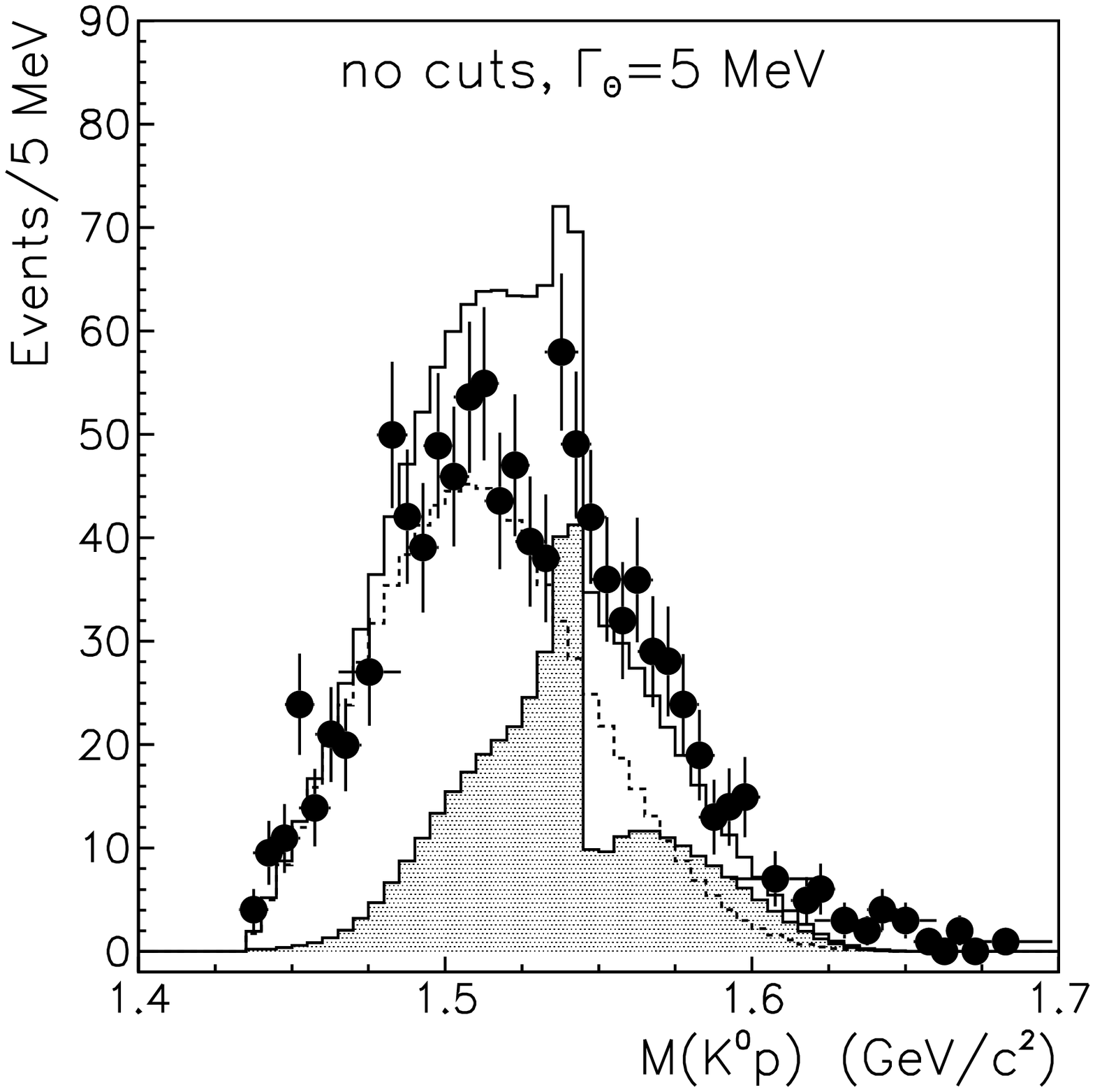,width=9.3cm,height=8.3cm}  &
\hspace*{-10mm}\psfig{file=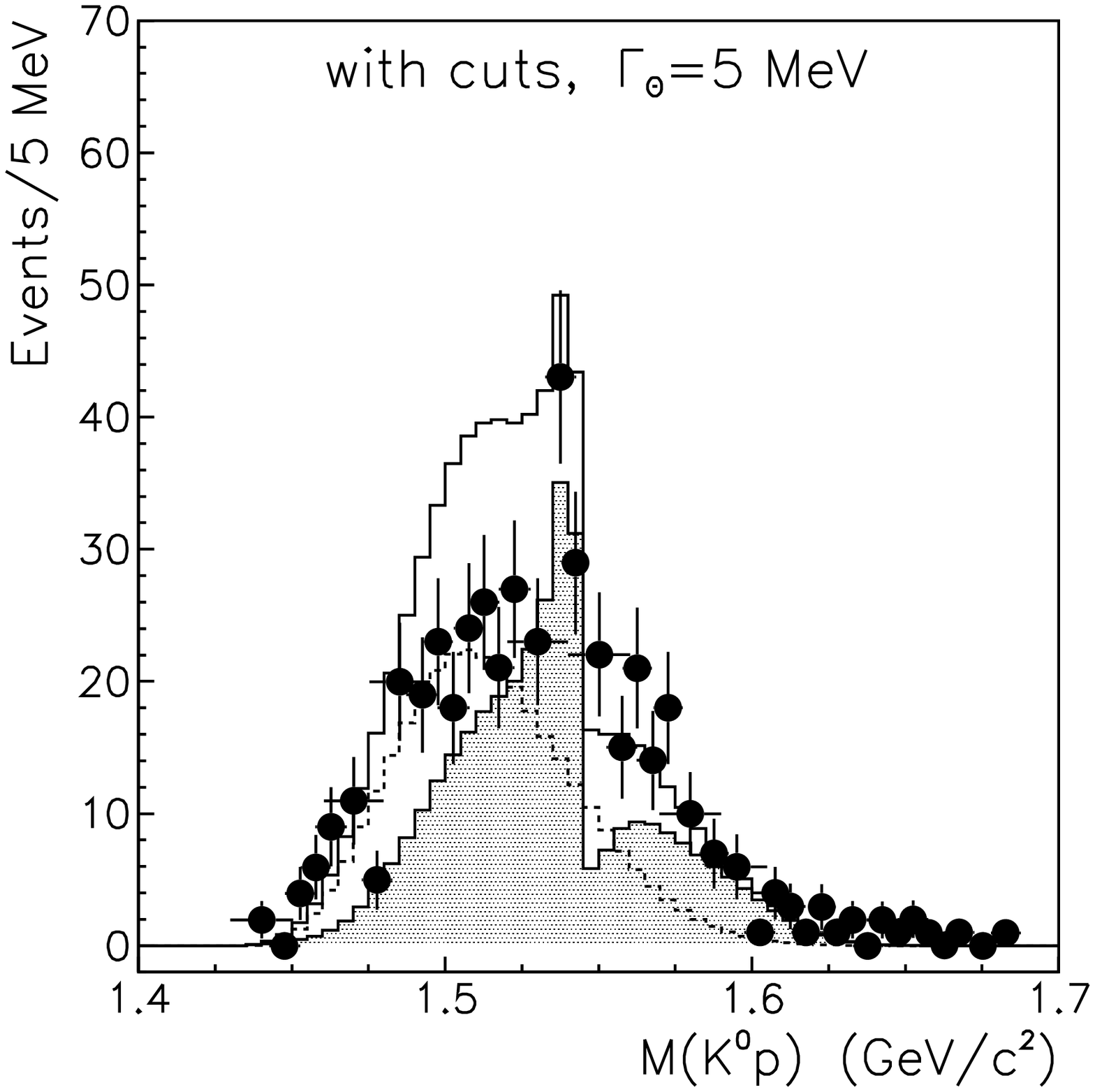,width=9.3cm,height=8.3cm} \\
\psfig{file=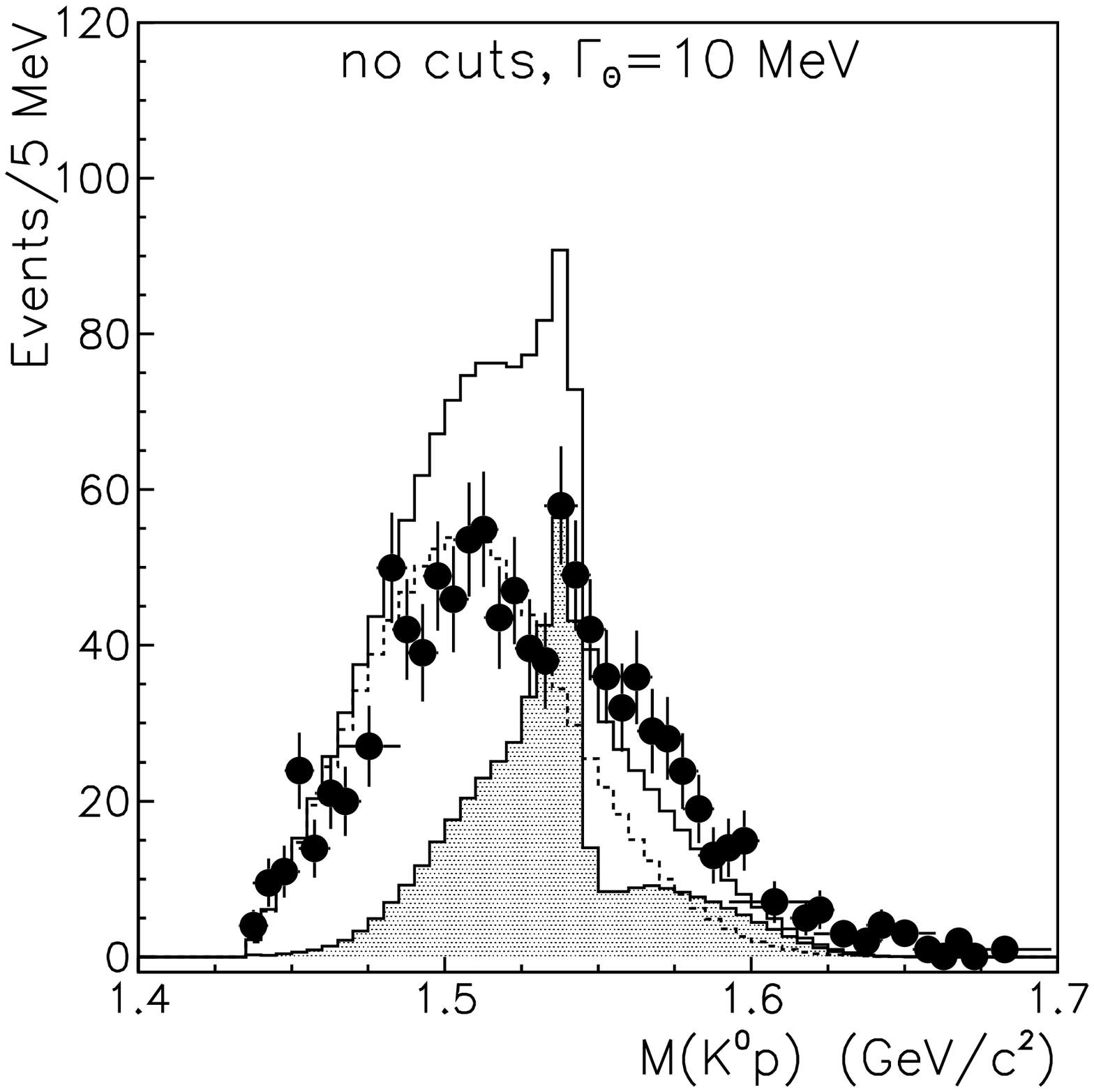,width=9.3cm,height=8.3cm}  &
\hspace*{-10mm}\psfig{file=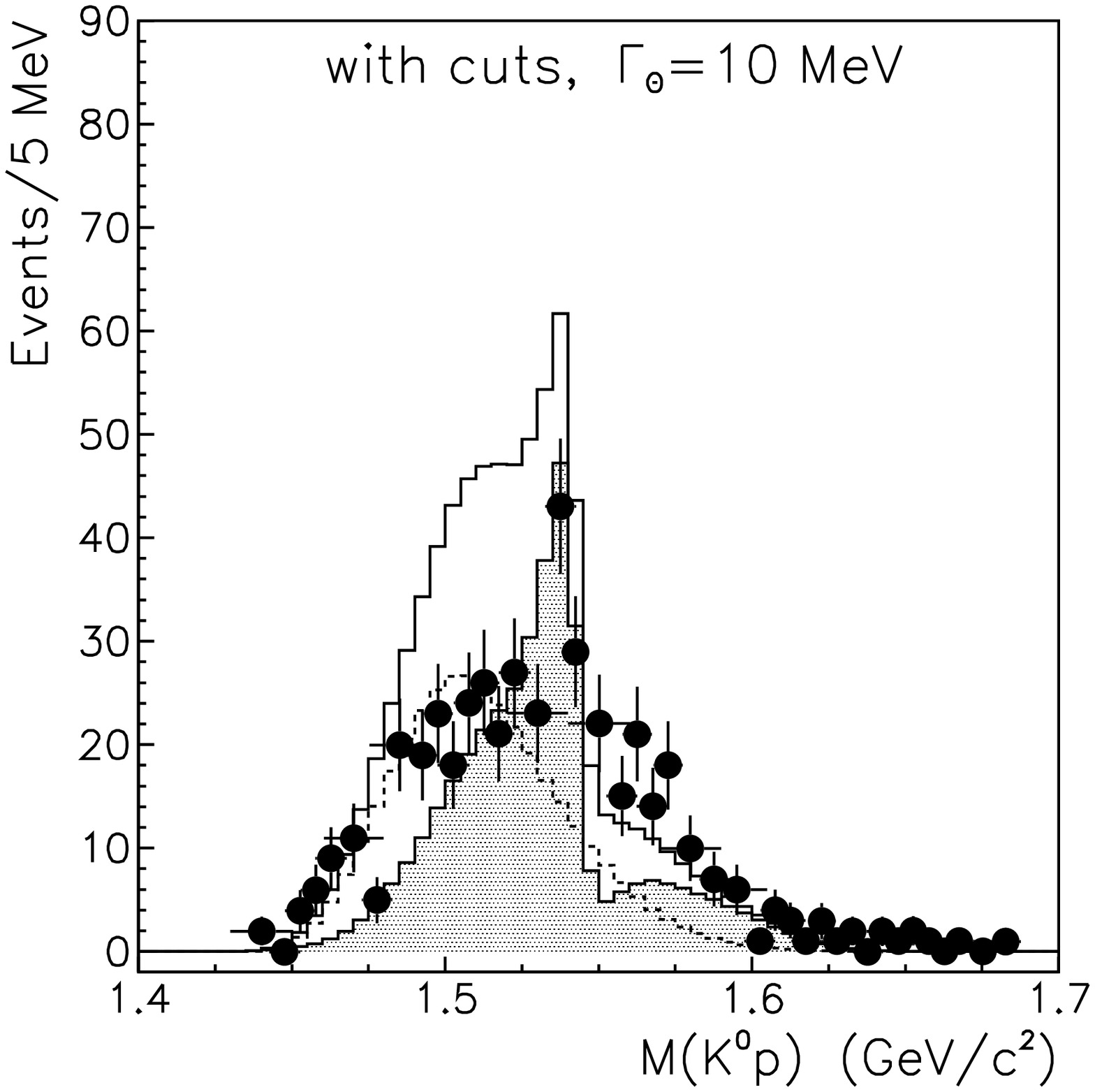,width=9.3cm,height=8.3cm} 
\end{tabular}
\vspace*{-4mm}
\caption{The same as in Fig.~\ref{diana2} but with the inclusion
of a $\Theta^+$ resonance with positive parity
and a width of 1~MeV (top), 5~MeV (middle) and 10~MeV (bottom).
The results are shown without and with kinematical cuts.}
\label{diana5}
\end{figure*}

Fig.~\ref{diana2} shows that the calculations based on Eq.~(\ref{react1})
substantially underestimate the data at low masses. This discrepancy
can be attributed to the rescattering of the protons after the
$K^+n{\to}K^0p$ reaction in nuclear medium.

The elastic $pN{\to}pN$ cross section on a bound nucleon is 
evaluated by 
\begin{eqnarray}
\frac{d^3\sigma}{dp d\Omega}{=}\!\!\!
\int \!\!\Phi (p_p) dp_p \!\!
\int \!\!P(E_q,q) |T_{NN}(s_{pN},\Omega)|^2 dE_qd^3q,\,\,\,\,\,\,\,\,
\end{eqnarray}
where $\Phi (p_p)$ is the proton spectrum from the reaction 
$K^+n{\to}K^0p$ given by Eq.~(\ref{react1})
and $T_{NN}$ is the nucleon-nucleon scattering amplitude.  The
squared invariant energy of the $pN$ interaction is defined 
analogous to Eq.~(\ref{skn}).

The dashed histograms in Fig.~\ref{diana2} show the $K^0p$
invariant mass spectrum calculated with taking into account the 
rescattering of the proton after the direct $K^+n{\to}K^0p$ reaction.
This spectrum reproduces the experimental results at low masses
of the $K^0p$ system rather well. 
The sum of the direct and rescattering processes is shown by the 
solid histograms in Fig.~\ref{diana2} and is evidently in quite good
agreement with the DIANA data~\cite{Dolgolenko}. 
It is important that our calculations
describe the data before and after the kinematical cuts performed in
Ref.~\cite{Dolgolenko}. The intention of those experimental cuts was 
to reduce the background from the rescattering of the final particles 
in the nuclear medium and our calculations explicitely confirm that
this goal is indeed achieved by these cuts. 
Furthermore, our results illustrate that the enhancement at the
$K^0p$ effective mass of around 1539~MeV is not produced 
by the kinematical cuts.

\section{Discussion of the $\Theta^+$ contribution}

The next step is to scrutinize whether the enhancement at the $K^0p$ mass of
around $M(K^0p){=}1539{\pm}2$~MeV is indeed caused by the $\Theta^+$ resonance 
contribution, which in our calculations enters via the $K^+n{\to}K^0p$
amplitude. For that we evaluate the $K^0p$ invariant mass spectra
for the reaction $K^+Xe{\to}K^0pX$ employing different widths of the
$\Theta^+$ resonance and considering positive as well as negative parity
for the $\Theta^+$. We always compare our results with the 
experimental information without and with kinematical cuts. In 
these calculations we use the same normalization to the data 
as adopted earlier in the considerations without the $\Theta^+$ contribution.

Fig.~\ref{diana5} shows the results for the $\Theta^+$ resonance with 
positive parity
and for a width of 1~MeV (top), 5~MeV (middle), and 10~MeV (bottom), in order. 
The contribution from the direct $K^+n{\to}K^0p$ process
is shown by the hatched histogram. It clearly reflects the 
particular bump-dip structure of the elementary $KN$ amplitude that we
discussed in the section II.  Following the experimental analysis
we use a binning of 5~MeV for the $K^0p$ mass spectrum, i.e. the result
of the calculation is integrated over an interval of 5~MeV.

Evidently, the model results for a width of 1~MeV
-- without and with kinematical cuts -- are in 
rather good agreement with the experimental data.
In the calculations for a $\Theta^+$ width of 5~MeV 
(middle section of Fig.~\ref{diana5})
the structure from the $\Theta^+$ resonance 
contribution becomes more pronounced and the enhancement at 
1539~MeV is described substantially better. However, the 
theoretical prediction misses the experimental points for 
invariant masses slightly larger than $M(K^0p){=}$1539~MeV, 
due to the particular structure of the $K^+n{\to}K^0p$ amplitude
mentioned before. With regard to the data after the
the kinematical cuts are in strong disagreement with the experiment.
With regard to the data after the 
kinematical cuts the inclusion of a $\Theta^+$ resonance with 
$\Gamma_{\Theta^+}$ = 5 MeV leads to a large shoulder at invariant masses 
just below the resonance energy which is not present in the experiment.

The bottom section of 
Fig.~\ref{diana5} presents our results obtained for a  
$\Theta^+$ resonance width of 10~MeV. Now the disagreement between 
our calculations and the DIANA experiment is significantly stronger.
For example, after the kinematical cuts the enhancement due to 
the $\Theta^+$ contribution overestimates the measured signal by a 
factor of about 1.6. Moreover, the predictions at low masses,
i.e. below 1539~MeV, significantly overestimate the measurements.
Clearly the DIANA data do not support such a large width of the 
$\Theta^+$ resonance. Note that the width of 10~MeV
corresponds roughly to the upper limit for the $\Theta^+$ width
of 9~MeV that was claimed by the DIANA collaboration \cite{Dolgolenko}.

\begin{figure*}
\begin{tabular}{cc}
\vspace*{-8mm}\psfig{file=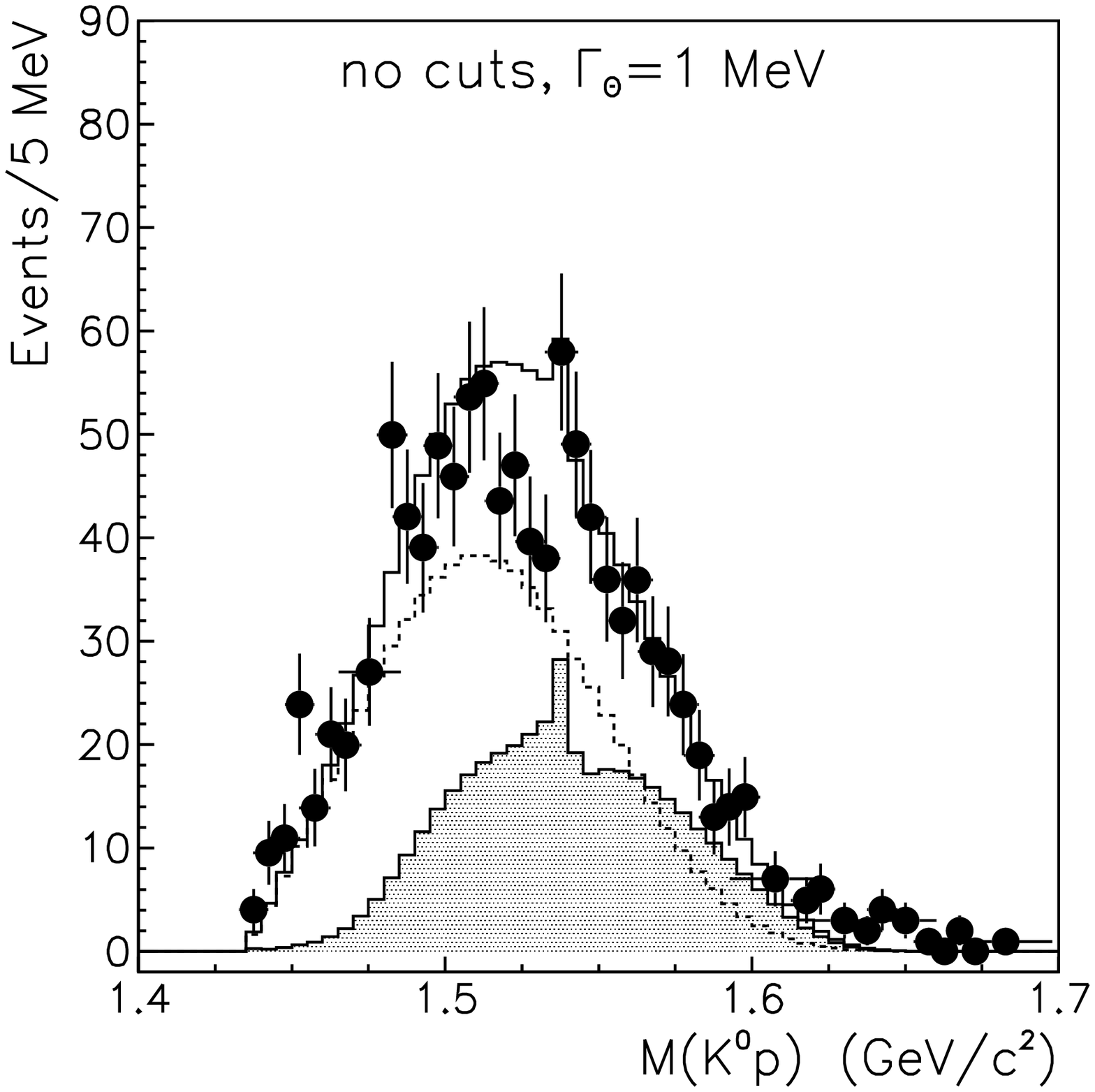,width=9.3cm,height=8.3cm} &
\hspace*{-10mm}\psfig{file=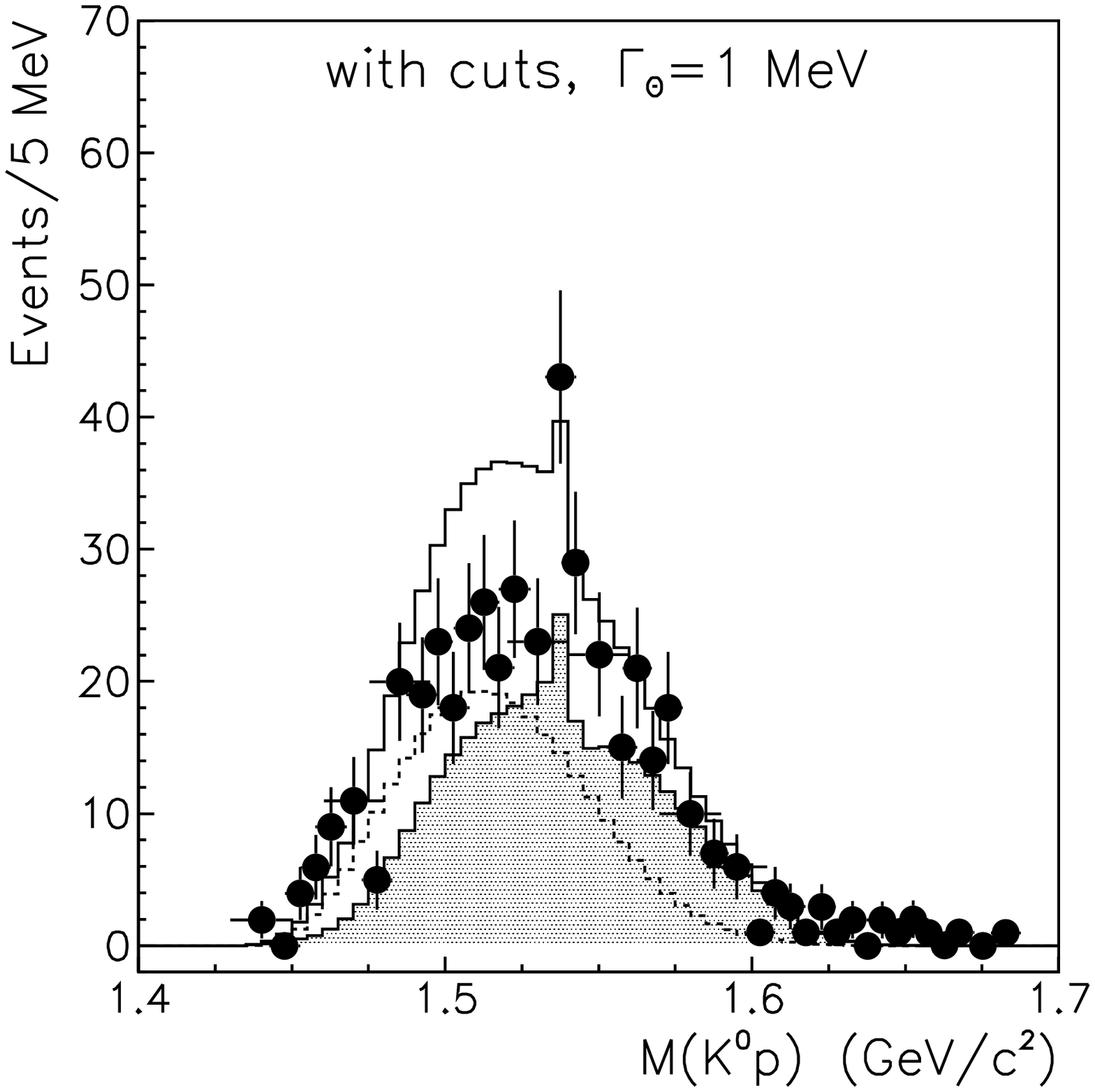,width=9.3cm,height=8.3cm} \\
\vspace*{-8mm}\psfig{file=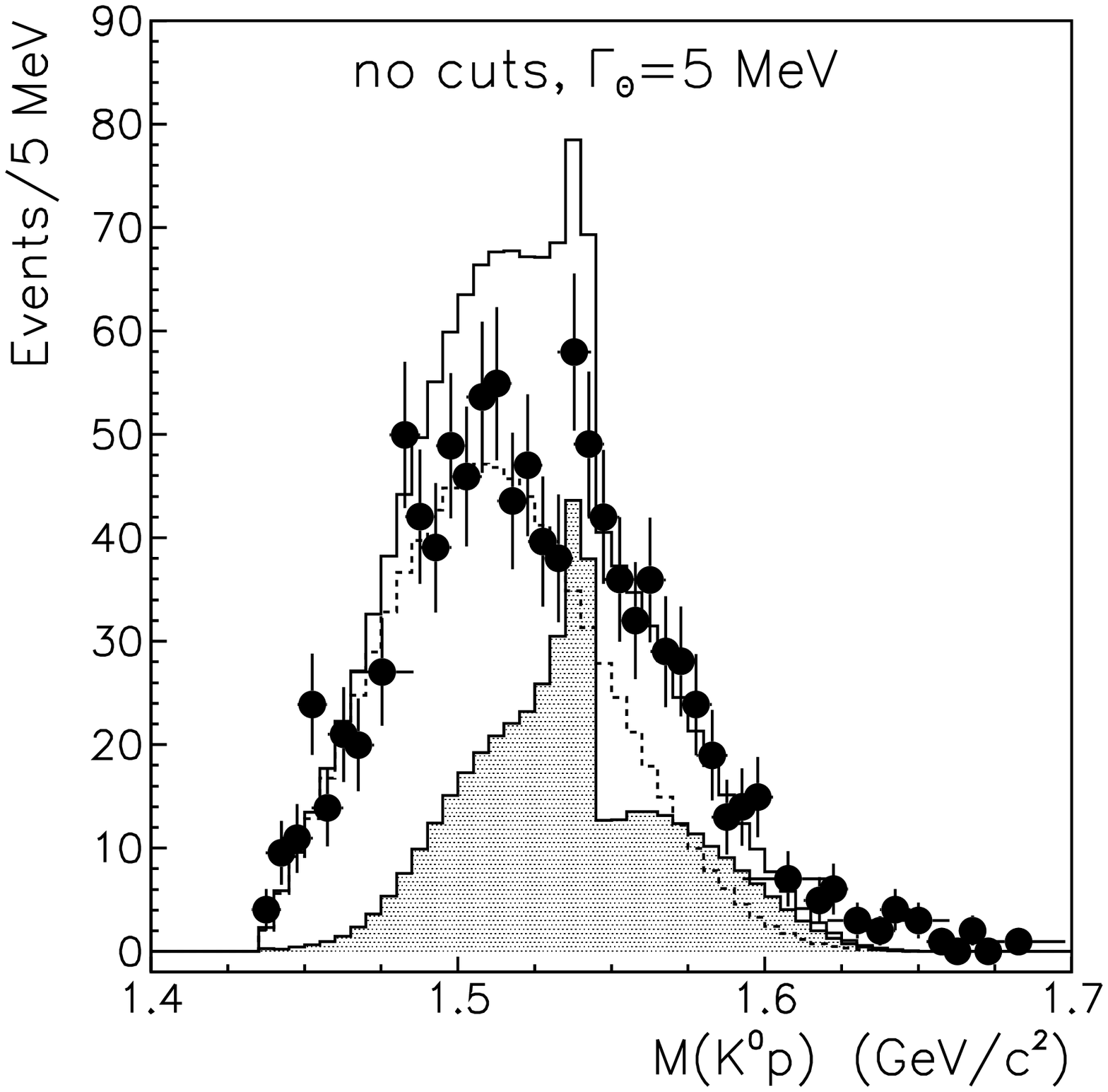,width=9.3cm,height=8.3cm}  &
\hspace*{-10mm}\psfig{file=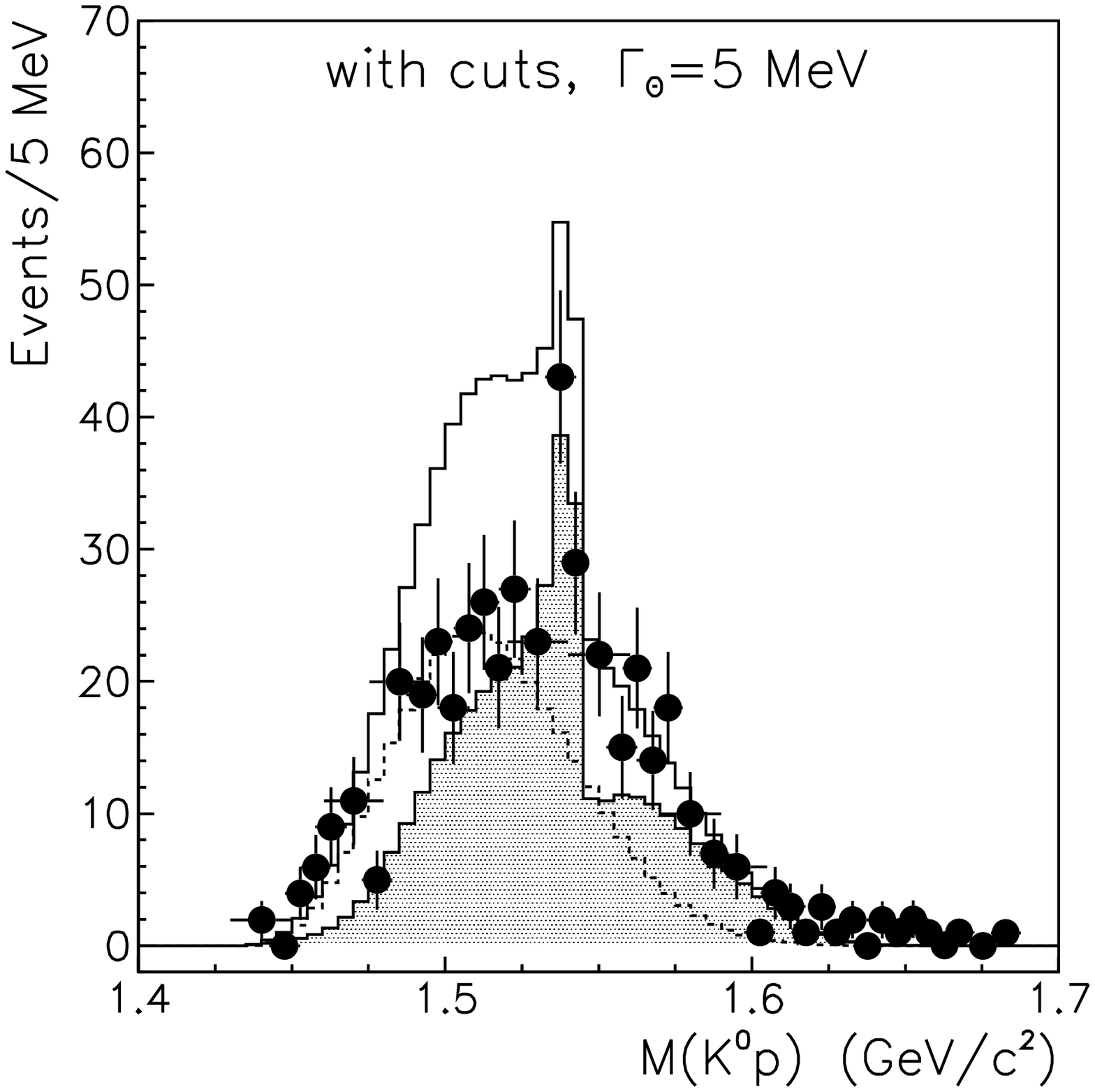,width=9.3cm,height=8.3cm} \\
\psfig{file=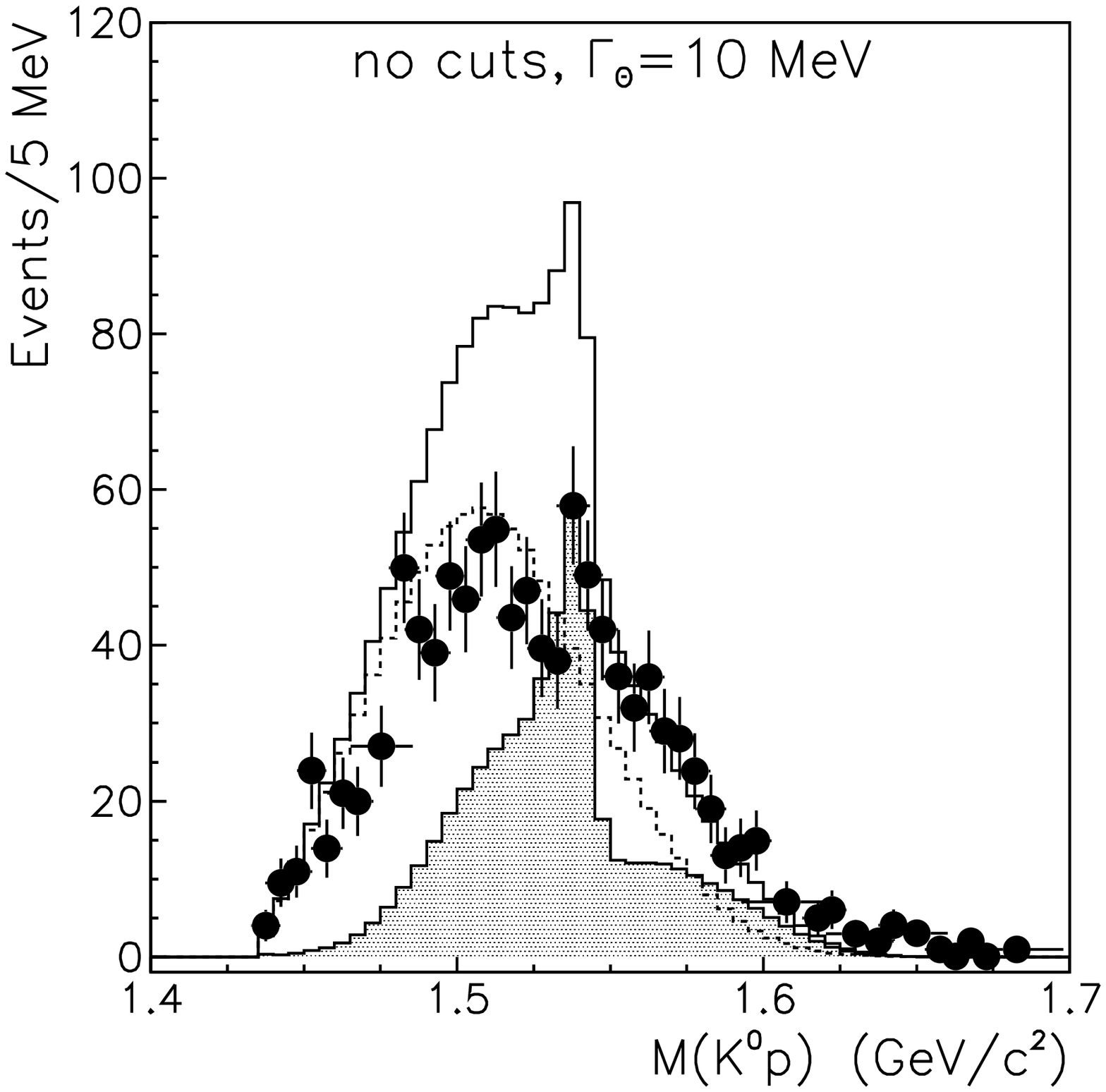,width=9.3cm,height=8.3cm}  &
\hspace*{-10mm}\psfig{file=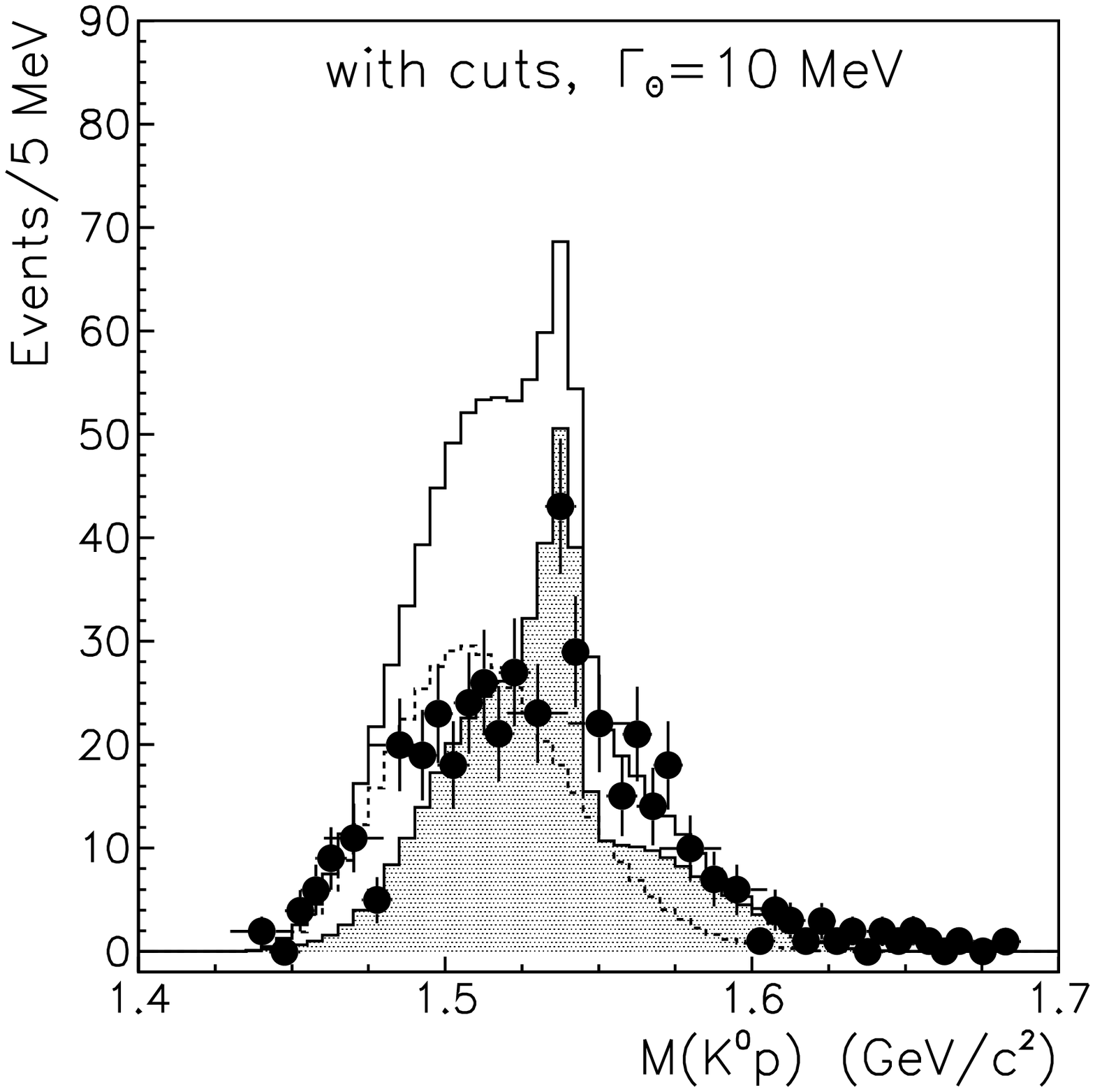,width=9.3cm,height=8.3cm} 
\end{tabular}
\vspace*{-4mm}
\caption{The same as in Fig.~\ref{diana2} but with the inclusion
of a $\Theta^+$ resonance with negative parity
and  a width of 1~MeV (top), 5~MeV (middle) and 10~MeV (bottom).
The results are shown without and with kinematical cuts.}
\label{diana11}
\end{figure*}

Results for the $\Theta^+$ assuming negative parity and again
for widths of 1~MeV, 5~MeV, and 10~MeV are presented in
Fig.~\ref{diana11}. Qualitatively the description of the DIANA
data is very similar to the case of positive parity. Specifically, 
there is no obvious evidence that the data favour a particular
parity of the $\Theta^+$. On a quantitative level one can see that
the negative parity case leads to a slightly more pronounced
shoulder for invariant masses $M(K^0p)$ below the resonance
energy, and that means to a stronger disagreement with the
data, whereas there is an improvement for invariant masses larger
than the resonance energy because the dip structure, which is not 
seen in the data, is now softened. 

In order to enable an easy quantitative comparison of the various 
scenarios considered we provide here also the corresponding 
$\chi^2$ values per data point, cf. Table 1. Evidently, the
calculation based on the original $KN$ model of the J\"ulich
group, without a $\Theta^+$ resonance, provides an excellent
description of the DIANA data before the kinematical cuts 
leading to a $\chi^2$/dof of 1.2. The $\chi^2$ deteriorates
significantly for the data where kinematical cuts were
applied. A closer inspection reveals that this deterioration
is mainly due to a depletion of the experimental 
invariant mass spectrum for large $M(K^0p)$ after the cuts
are applied. We do not have a ready explanation for this because,
given the nature (and the intention) of the cuts, we would
expect only a reduction for low values of $M(K^0p)$ -- as it
is the case in our model calculation. 

\begin{table}[htb]
\caption{\label{tab1} $\chi^2$/dof  evaluated by comparing
our calculations for different parity $\pi(\Theta^+)$ and for 
different $\Theta^+$ widths $\Gamma_{\Theta^+}$
with the experimental information on the $K^0p$ effective mass
spectrum for the reaction $K^+Xe{\to}K^0pX$ from the 
DIANA collaboration \protect\cite{Dolgolenko}.
Considered are the data before the kinematical cuts (44 
experimental points) as well as those where cuts were applied  
(39 points). The first line is the result without inclusion 
of a $\Theta^+$
resonance. 
}
\begin{center}
\begin{tabular}{||c|c|c|c||}
\hline
$\pi(\Theta^+)$ & $\Gamma_{\Theta^+}$ (MeV) & no cuts  & with cuts\\
\hline
 no &  no  & 1.2 & 2.3 \\
\hline
+ & 1 & 1.4 & 2.7 \\
+ & 5 & 3.0 & 3.9  \\
+ & 10 & 6.7 & 7.4  \\
\hline
-- & 1 & 1.4 & 2.9 \\
-- & 5 & 3.6 & 5.2  \\
-- & 10 & 9.2 & 11.2  \\
\hline
\end{tabular}
\end{center}
\end{table}

Adding a $\Theta^+$ resonance causes an increase rather than a 
decrease in the $\chi^2$ value
for all considered scenarios, cf. Table 1. This increase is 
clearly marginal for a $\Theta^+$ with a width of 1~MeV, but one
would have certainly expected an improvement in the $\chi^2$
once the structure seen in the DIANA data can be reproduced. 
However, looking at the DIANA results one can understand what happens. 
Besides the narrow structure associated with the $\Theta^+$
the data show a fall-off just left of it. The fall-off is very 
pronounced prior to the cuts but still visible even after 
the kinematical cuts were applied, cf. Fig.~\ref{diana2}.
That feature is well reproduced in the model calculation
without a $\Theta^+$. Adding the $\Theta^+$ generates the
resonance structure and brings the results closer to the
corresponding data point. But at the same time the dip
is filled up and a broad shoulder developes, as can
be seen in Fig.~\ref{diana5}, and therefore the agreement
with the data worses in this range of the $K^0p$ effective
mass. This feature is unavoidable in a microscropic 
calculation like ours because the background part of the
$KN$ amplitude is strongly constrained by the $KN$ data
and an added resonance structure can only interfere in
a particular way with the background as discussed in
Section II. Since the signal of the actual resonance
is strongly reduced as compared to the elementary $KN$ cross
section (cf. Fig.~\ref{knsig}), specifically for a
narrow width, because the calculated results are averaged 
over bins of 5 MeV -- as is done in the experiment --
it is practically impossible to reproduce the narrow
resonance structure and the close-by narrow dip at the
same time. This is the reason why the $\chi^2$ does
not improve even if a narrow resonance is considered. 

For larger widths the dip that develops to the right
side of the resonance structure, and which likewise is
a consequence of the consistent treatment of 
the background and resonance part of the elementary
$KN$ amplitude in our calculation, becomes more
pronounced and causes a further and rapid increase in 
the $\chi^2$/dof.

\section{Summary}

We have performed a detailed and microscopic
calculation of the $K^0p$ effective mass spectrum for the
charge-exchange reaction $K^+Xe \to K^0pX$. An enhancement
in this quantity at an effective mass close to 1540 MeV,
found by the DIANA collaboration \cite{Dolgolenko}, 
is seen as an evidence for the existence of an exotic $S=1$
($\Theta^+$) baryon resonance. 
 
The main ingredient of
our calculation is an elementary $KN$ scattering amplitude
that is taken from the meson-exchange model developed by
the J\"ulich group \cite{Juel1,Juel2}. The role and
significance of the $\Theta^+$ for a quantitative
description of the DIANA data is investigated by
employing variants of the J\"ulich $KN$ model that 
include a $\Theta^+$(1540) resonance, assuming spin 1/2.  
We considered positive as well as negative parity for the
$\Theta^+$ resonance and also a range of values for
the width of the resonance. 

Since the DIANA collaboration published results 
of the measured $K^0p$ mass spectrum prior and after 
kinematical cuts we could simulate the cuts
in our calculation and, specifically, we could 
compare to the data with and without kinematical cuts in order to
control the contribution from the direct $K^+n{\to}K^0p$ reaction
in the $Xe$ nucleus and the rescattering processes.
Thereby we could confirm that the broader structure seen in their
``raw'' data around the effective mass of 1500 MeV is indeed largely due 
to rescattering processes of the proton after the direct 
reaction $K^+n\to K^0p$. We could also rule out that the 
sharp structure seen around the $K^0p$ effective mass of 
$1539{\pm}2$ MeV is an artefact of the applied kinematical cuts. 

We found that the $K^+Xe{\to}K^0pX$ calculations without
$\Theta^+$ contribution as well as the results obtained with a
$\Theta^+$ width of 1~MeV are in comparably good agreement with 
the DIANA results. Any spin-1/2 resonance with a larger width can  
definitely be excluded. But we must say that even a very narrow 
structure is only marginally supported by the data.
In light of our results, the high rating of the $\Theta^+$ 
resonance in the newest version of the PDG tables~\cite{lastPLB} 
appears too optimistic. It is evident that more dedicated experiments, 
employing both strong (e.g. at COSY-TOF) and electromagnetic probes, are
mandatory to establish this exotic state.
 
\begin{acknowledgement}
We would like to thank A.G. Dolgolenko for useful discussions.
\end{acknowledgement}

\end{document}